\documentstyle[aps,eqsecnum]{revtex}
\begin{document}
\draft
\preprint{WUGRAV 95-15}
\title{Head-on collision of compact objects in general relativity:\\
       Comparison of post-Newtonian and perturbation approaches}
\author{Liliana E.~Simone, Eric Poisson, and Clifford M.~Will}
\address{McDonnell Center for the Space Sciences,
         Department of Physics, Washington University,
         St.~Louis, Missouri 63130}
\date{\today}
\maketitle
\begin{abstract}
The gravitational-wave energy flux produced during the head-on infall and
collision of two compact objects is calculated using two approaches:
(i) a post-Newtonian method, carried to second post-Newtonian
order beyond the quadrupole formula, valid for systems of arbitrary
masses; and (ii) a black-hole perturbation method, valid for a test-body
falling radially toward a black hole. In the test-body case, the
methods are compared.  The post-Newtonian method is shown to converge
to the ``exact'' perturbation result more slowly than expected
{\it a priori\/}.  A surprisingly good approximation to the energy
radiated during the infall phase, as calculated by perturbation
theory, is found to be given by a Newtonian, or quadrupole,
approximation combined with the exact test-body equations of motion
in the Schwarzschild spacetime.
\end{abstract}
\pacs{PACS numbers: 04.25.Nx, 04.30.-w, 97.60.Jd, 97.60.Lf}
\widetext
\section{Introduction and summary}

The head-on collision of two bodies in general relativity has long
been a test-bed for various approaches to solving Einstein's equations
in dynamical situations, despite its vanishingly small astrophysical
probability.  Because of the axial symmetry of the situation, the problem
simplifies dramatically, yet can still retain features of potential
astrophysical importance, such as strong curvature, fast motions, and
the emission of gravitational radiation to infinity.  The first
attempt to solve this problem in a fully general relativistic manner
was the 1971 calculation of perturbations of the Schwarzschild
black-hole spacetime caused by a test particle in radial infall from
infinity \cite{DRPP}, using the recently developed Zerilli equation
for black-hole perturbations \cite{Zerilli}.  The main result of
this calculation was a formula for the total gravitational-wave
energy emitted, $\Delta E =0.0104 \mu c^2(\mu /m)$, where
$\mu$ is the mass of the test body, and $m$ the mass of the black
hole ($\mu \ll m$). Subsequently, these perturbation methods were
extended to handle more general test-body orbits around black holes
\cite{DetweilerSzedenits,SasakiNakamura,KojimaNakamura}.

The head-on collision of two black holes has also been the starting
point in attempts to solve general relativistic, dynamical problems by
purely numerical techniques.  The pioneering computation of the
head-on-collision of two equal-mass black holes in the middle 1970s
by Smarr and Eppley \cite{SmarrEppley} stood for almost 15 years as
the state of the art, until recent technical advances together with
the use of supercomputers resulted in a substantial improvement in
accuracy and reliability \cite{Seidel}. Generalizing these computations
to the case of black holes in circular orbits is sufficiently difficult
that it is viewed as a ``grand challenge.''

Another method for studying the interaction of compact objects and the
emission of gravitational radiation is the post-Newtonian
approximation.  Although this method can be used for bodies of arbitrary
masses in arbitrary orbits, it breaks down when the gravitational
fields become strong and the velocities become comparable to the speed
of light.

The ultimate goal toward which these calculations are pointed is
the solution of the inspiral and coalescence of two compact objects
(black holes or neutron stars), and the calculation to high accuracy of
the emitted gravitational waveform.  Such inspiralling systems are
believed to be the most promising sources for detection of gravitational
waves by kilometer-scale laser interferometric gravitational observatories
(LIGO \cite{LIGO} in the US, VIRGO \cite{VIRGO} in Europe).  Furthermore,
it has been shown that very accurate gravitational-waveform templates
will be needed in order to infer characteristics of the sources from
the data, using matched filtering techniques \cite{Thorne1987}.
It is also widely believed that all three theoretical techniques will
play a role in this effort: post-Newtonian techniques for describing
much of the inspiral phase (until the system becomes too relativistic),
numerical techniques for the late inspiral and coalescence, and test-body
perturbation techniques for the special case of a small mass inspiralling
into a very massive black hole.  Consequently, it is important to
explore carefully the regions of validity and the accuracy of these
techniques, especially in regimes where they overlap.

In this paper we study the gravitational radiation emitted during the
head-on collision of two compact objects. In particular, we
compute the gravitational-wave luminosity as a function of their
relative separation. We use two techniques: (i) the post-Newtonian
approximation, carried to second order [order $(v/c)^4 \sim
(Gm/rc^2)^2$] beyond the quadrupole approximation; and (ii) the
test-body perturbation technique. Specializing the post-Newtonian
method to the case of a test-body falling from infinity toward a
massive body, we compare the two approaches.
Our main findings are these:

(1) The post-Newtonian approximation begins to fail when the bodies
are separated by about $10Gm/c^2$ (in harmonic coordinates),
where $m$ is the total mass of the
system. By failure, we mean that post-Newtonian corrections to the
energy flux become comparable to the leading-order, quadrupole term.

(2) The post-Newtonian approximation converges very slowly: at a
separation of $100Gm/c^2$, the first and second post-Newtonian
corrections to the luminosity are respectively 5 and 40 times
their expected sizes of $Gm/rc^2 \simeq 10^{-2}$, and
$(Gm/rc^2)^2 \simeq 10^{-4}$, relative to the leading-order term.

(3)  The test-body approach shows that for separations larger
than $4Gm/c^2$ (in Schwarzschild coordinates), the radiation
emitted is dominated by bremsstrahlung radiation generated by
the infalling particle. Thereafter, the radiation is dominated by
black-hole quasi-normal ringing.  The bremsstrahlung radiation
contributes only approximately 3 percent of the total
energy emitted.

(4)  At separations greater than about $10Gm/c^2$, the perturbation results
can actually be well approximated by a hybrid ``quadrupole'' formula for the
luminosity, in which the exact Schwarzschild test-body equations of
motion are used to evaluate the three time derivatives of the source
quadrupole moment.  The post-Newtonian values differ from the perturbation
results by 20\% at a separation of $45Gm/c^2$, 10\% at $75Gm/c^2$, and
1\% at $400Gm/c^2$, reflecting a very slow convergence.  The slow
convergence of the post-Newtonian approximation has also been noted in
the case of circular test-body orbits
\cite{CFPS,TagoshiNakamura,TagoshiSasaki,Poisson}.

The organization of this paper is as follows: Section II discusses the
head-on collision of bodies of arbitrary masses to second
post-Newtonian (2PN) order beyond the quadrupole approximation,
including the effects of gravitational-wave tails.  Section III
discusses the perturbation approach, and presents the energy flux for
test-body radial infall as a function of
retarded time, showing both the bremsstrahlung radiation
and the dominant quasi-normal ringing radiation.  In section IV we
compare the two approaches; to do so requires a careful mapping between
the post-Newtonian retarded source variables, expressed in harmonic
coordinates, and the retarded Schwarzschild coordinates of the
perturbation approach. Section V discusses the accuracy of the
post-Newtonian approach, and derives the hybrid ``quadrupole''
formula for the luminosity. In an Appendix we discuss a remarkable
series of cancellations that occur in the post-Newtonian method
for the radial infall from infinity.

\section{Head-on collision of compact objects to
second post-Newtonian order}
\subsection{Gravitational wave generation to
second post-Newtonian order}

This first subsection summarizes from the literature the
relevant equations for gravitational wave generation
by a slowly moving system of two arbitrary, non spinning,
masses \cite{Will}.

We describe the motion of the two bodies in terms of their
relative distance. In a harmonic coordinate system
whose origin is at the center of mass,
the relative position vector
is $\bbox{x}\equiv \bbox{x}_1-\bbox{x}_2$,
where $\bbox{x}_1$ and $\bbox{x}_2$ are the positions
of each body. If the masses are denoted $m_1$ and $m_2$,
we define the total mass $m\equiv m_1+m_2$, the reduced mass
$\mu\equiv m_1\,m_2/m$, $\eta\equiv\mu/m$, and
$\delta m\equiv m_1-m_2$. Also,
$r\equiv |\bbox{x}|$ and $\bbox{v}\equiv\dot{\bbox{x}}$,
where a dot denotes differentiation with respect to time.

The luminosity (or energy flux) associated with the
gravitational waves can be expressed in terms of the
mass and current multipole moments of the radiation field
\cite{thorne}. We need only a few terms for an expression
accurate through second post-Newtonian (2PN) order beyond
the leading-order, quadrupole-formula expression:
\begin{equation}
\dot{E}(T) = \frac{1}{5}
\left(
{}^{(3)}{{\rm I}\!\!\!{\scriptscriptstyle{{}^{-}}}}^{ij}
\right)^2 +\frac{1}{189} \left(
{}^{(4)}{{\rm I}\!\!\!{\scriptscriptstyle{{}^{-}}}}^{ijk}
\right)^2 +\frac{1}{9072} \left(
{}^{(5)}{{\rm I}\!\!\!{\scriptscriptstyle{{}^{-}}}}^{ijkl}
\right)^2 +\cdots +\frac{16}{45} \left(
{}^{(3)}{{\rm J}\!\!\!{\scriptscriptstyle{{}^{-}}}}^{ij}
\right)^2 +\frac{1}{84} \left(
{}^{(4)}{{\rm J}\!\!\!{\scriptscriptstyle{{}^{-}}}}^{ijk}
\right)^2 +\cdots.
\label{lumi}
\end{equation}
Here,
${{\rm I}\!\!\!{\scriptscriptstyle{{}^{-}}}}^{\{i_n\}}$ and
${{\rm J}\!\!\!{\scriptscriptstyle{{}^{-}}}}^{\{i_n\}}$ are
respectively
the symmetric-trace-free (STF) mass and current multipole moments
of the radiation field, and the antescript denotes the number of
derivatives taken with respect to time. The luminosity is
expressed as a function of time $T$, which represents proper
time as measured by static observers at large distances.
The total energy radiated, from initial time $T_i$ to
final time $T_f$, is given by
\begin{equation}
\Delta E= \int_{T_i}^{T_f} \dot{E}(T)\, dT.
\label{mass}
\end{equation}

The radiative multipole moments in Eq.~(\ref{lumi}) can be
related to the multipole moments of the source. These are
expanded in powers of $v^2 \sim m/r$ to the order required
for an expression for the luminosity accurate through 2PN order.
They can be expressed as
\begin{mathletters}
\label{massmul}
\begin{eqnarray}
{{\rm I}\!\!\!{\scriptscriptstyle{{}^{-}}}}^{ij} &=&
\mu\Biggl\{\left[1+\frac{29}{42}\,(1-3\eta)\,v^2
 -\frac{1}{7}\,(5-8\eta)\left(\frac{m}{r}\right)\right] x^i\,x^j
 -\frac{4}{7}\,(1-3\eta)\,r\,\dot r\,x^i\,v^j
 +\frac{11}{21}\,(1-3\eta)\,r^2\,v^i\,v^j \nonumber\\
&& \mbox{} +x^i\,x^j
\biggl[\frac{1}{504}\left(253-1835\,\eta +3545\,\eta^2\right) v^4
+\frac{1}{756}\left(2021-5947\,\eta-4883\,\eta^2\right) v^2
\left(\frac{m}{r}\right) \nonumber\\
&& \mbox{}
-\frac{1}{252}\left(355+1906\,\eta-337\,\eta^2\right)
\left(\frac{m}{r}\right)^2
-\frac{1}{756}\left(131-907\,\eta+1273\,\eta^2\right)
{\dot r}^2\left(\frac{m}{r}\right)\biggr] \nonumber\\
&& \mbox{} + r^2\,v^i\,v^j\biggl[\frac{1}{189}
   \left(742-335\,\eta-985\,\eta^2\right)
   \left(\frac{m}{r}\right)
   +\frac{1}{126}\left(41-337\,\eta+733\,\eta^2\right) v^2
   +\frac{5}{63}\left(1-5\eta+5\eta^2\right){\dot r}^2\biggr]
 \nonumber\\
&& \mbox{} - r\,\dot r\,v^i\,x^j
   \biggl[\frac{1}{378}\left(1085-4057\,\eta-1463\,\eta^2\right)
     \left(\frac{m}{r}\right)
   +\frac{1}{63}\left(26-202\,\eta+418\,\eta^2\right) v^2\biggr]
 \Biggr\}_{{\rm STF}} + \,\,
{{\rm I}\!\!\!{\scriptscriptstyle{{}^{-}}}}^{ij}_{\rm Tail},
\label{massmula}    \\
{{\rm I}\!\!\!{\scriptscriptstyle{{}^{-}}}}^{ijk} & = &
-\mu\frac{\delta m}{m}
 \Biggl\{ \left[1+\frac{1}{6}\,(5-19\,\eta)\,v^2
 -\frac{1}{6}\,(5-13\,\eta)
  \left(\frac{m}{r}\right)\right]\,x^i\,x^j\,x^k
 +(1-2\eta)(r^2\,v^i\,v^j\,x^k-r\,\dot r\,v^i\,x^j\,x^k)
  \Biggr\}_{{\rm STF}},
\label{massmulb} \\
{{\rm I}\!\!\!{\scriptscriptstyle{{}^{-}}}}^{ijkl} & = &
\mu\left(1-3\eta\right) (x^i\,x^j\,x^k\,x^l)_{{\rm STF}},
\label{massmulc} \\
{{\rm J}\!\!\!{\scriptscriptstyle{{}^{-}}}}^{ij} & = &
-\mu\frac{\delta m}{m}
 \Biggl\lgroup \varepsilon^{iab}\,
 \biggl\{ \left[1+\frac{1}{2}\,(1-5\eta)\,v^2
 +2\,(1+\eta)\left(\frac{m}{r}\right)\right]x^jx^av^b
\nonumber \\ & & \mbox{}
  +\frac{1}{28}\frac{d}{dt}\left[(1-2\eta)
  (3r^2v^j-r\dot rx^j)\,x^av^b\right]
\biggr\} \Biggr\rgroup_{{\rm STF}},
\label{currmuld} \\
{{\rm J}\!\!\!{\scriptscriptstyle{{}^{-}}}}^{ijk} & = &
\mu\left(1-3\eta\right)
(\varepsilon^{iab}\, x^a\,v^b\,x^j\,x^k\,x^l)_{{\rm STF}}.
\label{currmule}
\end{eqnarray}
\end{mathletters}

The radiation measured at time $T$ by an observer at a large
distance $R$ from the source is made up of two contributions.
The first can be thought of as radiation propagating directly
from the source; it depends on the state of the source at
retarded time $u \simeq T-R$ (see below for a more precise
expression). The second can be thought of as radiation scattered
by the spacetime curvature as it propagates away from the source.
This contribution, called the ``tail,'' is emitted by the source
at all times earlier than $u$. The tail term can also be decomposed
into multipole moments. The dominant term, and the only one needed
to evaluate $\dot{E}$ to 2PN order, is given by the following
integral over retarded time along the source's trajectory
${\cal C}(u')$ prior to retarded time $u$ \cite{blan-damour}:
\begin{equation}
{{\rm I}\!\!\!{\scriptscriptstyle{{}^{-}}}}^{ij}_{\rm Tail}(u)=
2m \int_{{\cal C}(u')}
{}^{(2)}{{\rm I}\!\!\!{\scriptscriptstyle{{}^{-}}}}^{ij}
\left(u-u'\right)
\left[\ln\left(\frac{u'}{2{\cal S}}\right)+
\frac{11}{12}\right]\,du' .
\label{gentail}
\end{equation}
Here, $u'$ parametrizes the trajectory up to the final point
$u'=0$. The constant ${{\cal S}}$ is an arbitrary scale parameter
which arises from the joining of two coordinate systems (see Sec.~IIC
for a discussion); we will show that our results do not depend on
the value of this parameter.

The last element needed to compute the energy flux are the
equations of motion. For any two-body problem the relative
acceleration $\bbox{a} = \bbox{a}_1 - \bbox{a}_2$ in a harmonic
coordinate system with origin at the center of mass, accurate to
2PN order, is given by \cite{linc-will}
\begin{equation}
\bbox{a}=-\left(\frac{m}{r^2}\right)
    \bigl[(1+A_1+A_2)\,\bbox{n} +(B_1+B_2)\,\bbox{v}\bigr],
\label{twobodyacc}
\end{equation}
where $\bbox{n} = \bbox{x}/r$, and
\begin{mathletters}
\label{accelterms}
\begin{eqnarray}
A_1&=&-2\,(2+\eta)\left(\frac{m}{r}\right)
	     +(1+3\eta)\,v^2-\frac{3}{2}\,\eta\,\dot r^2,   \\
A_2&=&\frac{3}{4}\,(12+29\,\eta)\left(\frac{m}{r}\right)^2
	     +\eta\,(3-4\eta)\,v^4
	     +\frac{15}{8}\,\eta\,(1-3\eta)\,\dot r^4
      -\frac{3}{2}\,\eta\,(3-4\eta)\,v^2\,\dot r^2
\nonumber \\ & & \mbox{}
   -\frac{1}{2}\,\eta\,(13-4\eta)
   \left(\frac{m}{r}\right) v^2
   -(2+25\,\eta+2\eta^2)
    \left(\frac{m}{r}\right)\dot r^2, \\
B_1&=&-2\,(2-\eta)\,\dot r,  \\
B_2&=&\frac{1}{2}\,\dot r\left[(4+41\,\eta+8\eta^2)
      \left(\frac{m}{r}\right)
      -\eta\,(15+4\eta)\,v^2+3\eta\,
       (3+2\eta)\,\dot r^2\right].
\end{eqnarray}
\end{mathletters}
We substitute Eq.~(\ref{twobodyacc})
to the required order whenever time derivatives of the velocity
appear in the various terms comprising Eq.~(\ref{lumi}).

Solving Eq.~(\ref{twobodyacc}) yields the trajectory
$\bbox{x}(t)$. As in Newtonian mechanics, solving this equation
is facilitated by making use of the first integrals: $E$ the energy,
and $\bbox{J}$ the angular momentum. Both of these quantities are
conserved to 2PN order. Below we will specialize to the case of a
head-on collision, for which $\bbox{J}=0$. We therefore only need
an expression for the energy, which is given by~\cite{kidder}
\begin{eqnarray}
E &=& \mu\biggl\{\frac{1}{2}\,v^2  -\left(\frac{m}{r}\right)
  + \frac{3}{8}(1-3\eta)\,v^4
  +\frac{1}{2}(3+\eta)\left(\frac{m}{r}\right)\,v^2
  +\frac{1}{2}\eta\left(\frac{m}{r}\right)\dot r^2
  +\frac{1}{2}\left(\frac{m}{r}\right)^2
  + \frac{5}{16}(1-7\eta+13\,\eta^2)\,v^6
\nonumber\\ & & \mbox{}
  +\frac{1}{8}(21-23\,\eta-27\,\eta^2)\left(\frac{m}{r}\right)\,v^4
  +\frac{1}{4}\eta\,(1-15\,\eta)\left(\frac{m}{r}\right)\,v^2\,\dot r^2
 - \frac{3}{8}\eta\,(1-3\eta)\left(\frac{m}{r}\right)\dot r^4
\nonumber\\ & & \mbox{}
  +\frac{1}{8}(14-55\,\eta+4\eta^2)\left(\frac{m}{r}\right)^2 v^2
+\frac{1}{8}(4+69\,\eta+12\,\eta^2)\left(\frac{m}{r}\right)^2\dot r^2
-\frac{1}{4}(2+15\,\eta)\left(\frac{m}{r}\right)^3\biggr\}.
\label{twobodyenergy}
\end{eqnarray}

\subsection{Head-on collision}

In this subsection we use the equations listed above to
calculate the gravitational-wave luminosity produced
during the head-on infall and collision of two bodies of
arbitrary masses. We consider two different situations. In
the first (A), the infall proceeds from rest at infinite
initial separation; in the second (B), the infall proceeds
from rest at finite initial separation.

There is only one direction of motion in this problem. We
therefore have $\bbox{x}=z\, \bbox{n}$,
$\bbox{v}=\dot{z}\,\bbox{n}$, $r=z$, and $v=\dot r
=\dot z$. Using this, Eqs.~(\ref{lumi}) and~(\ref{massmul})
simplify considerably. Moreover, since the
current moments are all proportional to $\bbox{J}=0$,
they all vanish. The mass multipole moments reduce to:
\begin{mathletters}
\label{multip}
\begin{eqnarray}
{{\rm I}\!\!\!{\scriptscriptstyle{{}^{-}}}}^{ij}
& = & \mu z^2\biggl[1+\frac{9}{14}\,(1-3\eta)\,\dot z^2
- \frac{1}{7}\,(5-8\eta) \left(\frac{m}{z}\right)
- \frac{1}{252} \left(355+1096\,\eta-337\,\eta^2\right)
    \left(\frac{m}{z}\right)^2
\nonumber\\
& & \mbox{}
+ \frac{1}{126} \left(448+289\,\eta-1195\,\eta^2\right)
 \left(\frac{m}{z}\right)\dot z^2
+ \frac{1}{168} \left(83-589\,\eta+1111\,\eta^2\right)
 \dot z^4 \biggr] \bar{\bbox{P}}^{(2)}+\,\,
{{\rm I}\!\!\!{\scriptscriptstyle{{}^{-}}}}^{ij}_{\rm Tail},
\label{multipa}       \\
{{\rm I}\!\!\!{\scriptscriptstyle{{}^{-}}}}^{ijk}
& = & -\mu z^3\,\frac{\delta m}{m}
  \left[1+\frac{1}{6}\,(5-19\eta)\,\dot z^2
  -\frac{1}{6}\,(5-13\eta)
  \left(\frac{m}{z}\right)\right] \bar{\bbox{P}}^{(3)},\\
{{\rm I}\!\!\!{\scriptscriptstyle{{}^{-}}}}^{ijkl}
& = & \mu z^4\left(1-3\eta\right)
    \,\bar{\bbox{P}}^{(4)}.
\label{multipc}
\end{eqnarray}
\end{mathletters}
Here, the $\bar{\bbox{P}}^{(q)}$'s are STF products
of a number $q$ of unit vectors, as given by
Pirani~\cite{pirani} and Thorne~\cite{thorne}:
\begin{mathletters}
\label{tensors}
\begin{eqnarray}
\bar{\bbox{P}}^{(2)} & \equiv & n^i n^j-\frac{1}{3}\,
   \delta^{ij}
   \equiv {\bbox{P}}^{(2)} -
   \frac{1}{3}\,{\bbox{D}},
   \label{tentwo}\\
\bar{\bbox{P}}^{(3)} & \equiv & n^i n^j n^k-\frac{1}{5}
    \left(\delta^{ij} n^k+\delta^{ik} n^j +
    \delta^{jk} n^i\right)
   \equiv {\bbox{P}}^{(3)} -
   \frac{3}{5}\left({\bbox{D}}
   \cdot{\bbox{P}}^{(1)}\right),
   \label{tenthr}\\
\bar{\bbox{P}}^{(4)} & \equiv & n^i n^j n^k n^l-\frac{1}{7}
    \left(\delta^{ij} n^k n^l+\delta^{ik} n^j n^l +
    \delta^{il} n^k n^j
    +\delta^{jk} n^i n^l+\delta^{jl} n^i n^k +
    \delta^{jk} n^i n^l\right)
\nonumber\\ & & \mbox{}
   +\frac{1}{35} \left(\delta^{ij}\delta^{kl}+
   \delta^{ik}\delta^{jl}
   +\delta^{il}\delta^{kj}\right)
    \equiv {\bbox{P}}^{(4)} -
     \frac{6}{7}\left({\bbox{D}}
     \cdot{\bbox{P}}^{(2)}\right)
   +\frac{3}{35}\left({\bbox{D}}\cdot
     {\bbox{D}}\right);
\label{tenfou}
\end{eqnarray}
\end{mathletters}
the ${\bbox{P}}^{(q)}$'s are symmetric products of a number
$q$ of unit vectors, and $\bbox{D}$ symbolically represents
the Kronecker delta. The compact notation to the right-hand-side
of Eqs.~(\ref{tensors}), with which we represent fully symmetrized
products of symmetric tensors by enclosing them within parentheses
[e.g., (${\bbox{D}}\cdot{\bbox{P}}^{(1)})$ or
(${\bbox{D}}\cdot{\bbox{D}}$)], corresponds to
expressions~(2.1) and~(2.2) in Ref.~\cite{thorne}.
Notice that
${{\rm I}\!\!\!{\scriptscriptstyle{{}^{-}}}}^{ij}_{\rm Tail}$
in Eq.~(\ref{multipa}) also contains a factor
$\bar{\bbox{P}}^{(2)}$; it arises via the term
${}^{(2)}{{\rm I}\!\!\!{\scriptscriptstyle{{}^{-}}}}^{ij}$
in the integrand of Eq.~(\ref{gentail}).

As the movement is directed along the $z$ axis,
the ${\bbox{P}}^{(q)}$'s have components along $z$ only
(e.g., ${\bbox{P}}^{(2)} = \delta^{iz}\delta^{jz}$).
The $\bar{\bbox{P}}^{(q)}$'s are therefore diagonal tensors,
constant in time, and they factor out of the time derivatives
in Eq.~(\ref{lumi}). As a result, we need only calculate
the squares of the $\bar{\bbox{P}}^{(q)}$'s,
\begin{equation}
\bar{\bbox{P}}^{(2)} \cdot \bar{\bbox{P}}^{(2)}
=\frac{2}{3},\qquad
\bar{\bbox{P}}^{(3)} \cdot \bar{\bbox{P}}^{(3)}
=\frac{2}{5},\qquad
\bar{\bbox{P}}^{(4)} \cdot \bar{\bbox{P}}^{(4)}
=\frac{8}{35},
\label{factors}
\end{equation}
and then insert them into the squares of
${}^{(n+1)}{{\rm I}
\!\!\!{\scriptscriptstyle{{}^{-}}}}^{i_1\cdots i_n}$
in Eq.~(\ref{lumi}).

We must now evaluate the time derivatives of the mass multipole
moments, substituting the expression for the acceleration
$\ddot z$ when needed. For a head-on collision,
Eq.~(\ref{twobodyacc}) reduces to
\begin{equation}
\ddot z = -\frac{m}{z^2}
\Bigg[ 1-2\,(2+\eta)\left(\frac{m}{z}\right)
-\left(3-\frac{7}{2}\eta\right)\dot z^2
+ \left(\frac{m}{z}\right)^2
\left(9+\frac{87}{4}\eta\right)
-\left(\frac{m}{z}\right)\,
\dot z^2\,\eta\,(11-4\eta)
-\frac{21}{8}\,\dot z^4\,\eta\,(\eta+1) \Bigg].
\label{motion}
\end{equation}
It is also helpful to make use of our expression
(\ref{twobodyenergy}) for the conserved energy in
order to express $\dot{z}$ as a function of $m/z$.
For a radial infall, we have
\begin{eqnarray}
E &=& \mu\Biggl[ \frac{1}{2}\,\dot z^2-
\left(\frac{m}{z}\right) +\frac{3}{8}\,(1-3\eta)\,\dot z^4
  +\frac{1}{2}\,(3+2\eta)\,\left(\frac{m}{z}\right)\dot z^2
  +\frac{1}{2}\left(\frac{m}{z}\right)^2
  +\frac{5}{16}\,(1-7\eta+13\,\eta^2)\,\dot z^6 \nonumber\\
&& \mbox{} +\frac{3}{8}\,(7-8\eta-16\,\eta^2)
   \left(\frac{m}{z}\right)\dot z^4
  +\frac{1}{4}\,(9+7\eta+8\eta^2)
   \left(\frac{m}{z}\right)^2\dot z^2
  -\frac{1}{4}\,(2+15\,\eta)
   \left(\frac{m}{z}\right)^3\Biggr].
\label{headonenergy}
\end{eqnarray}
Defining $z_0$ to be the initial separation (at which $\dot z=0$),
we can invert Eq.~(\ref{headonenergy}) for $\dot z(z,z_0)$.
For case (A), in which the initial separation is infinite,
$E(z) = E(\infty) = 0$, and we obtain
\begin{equation}
\dot z= -\Biggl\{\frac{2m}{z} \biggl[ 1-
5\Bigl(1-\frac{\eta}{2}\Bigr)
\Bigl(\frac{m}{z}\Bigr)
+\Bigl(13-\frac{81}{4}\eta+5\,\eta^2\Bigr)
\Bigl(\frac{m}{z}\Bigr)^2\biggr]\Biggr\}^{1/2}.
\label{veloinf}
\end{equation}
For case (B), in which the initial separation is finite,
we have
\begin{equation}
E(z) = E(z_0) = - \mu\left[
   \left(\frac{m}{z_0}\right) -
    \frac{1}{2}\left(\frac{m}{z_0}\right)^2
   + \frac{1}{2}\,\left(1+\frac{15}{2}\eta \right)\,
    \left(\frac{m}{z_0}\right)^3
    \right],
\end{equation}
and we obtain
\begin{eqnarray}
\dot z & = & -\Biggl\{2\,\left(\frac{m}{z}-
   \frac{m}{z_0}\right)\bigg[
   1-5\left(\frac{m}{z}\right)
   \left(1-\frac{\eta}{2}\right)
   +\left(\frac{m}{z_0}\right)
   \left(1-\frac{9}{2}\,\eta\right)
 +\left(\frac{m}{z}\right)^2
    \left(13-\frac{81}{4}\eta+5\,\eta^2\right)
\nonumber \\ & & \mbox{}
  +\left(\frac{m}{z}\right)\left(\frac{m}{z_0}\right)
  \left(5-\frac{173}{4}\eta+13\,\eta^2\right)
  +\left(\frac{m}{z_0}\right)^2
   \left(1-\frac{5}{4}\eta+8\,\eta^2\right)
   \bigg]\Biggr\}^{1/2}.
\label{velozo}
\end{eqnarray}
We can then insert Eqs.~(\ref{motion}), (\ref{veloinf}),
and Eq.~(\ref{velozo}) into the time derivatives of
the mass moments, Eqs.~(\ref{multip}), and calculate the
gravitational-wave luminosity in terms
of $m/z$ and $m/z_0$.

We discuss in the Appendix an interesting cancellation of
terms which occurs in the case of an infall from infinity.

\subsection{Tail terms}

We now turn to the evaluation of the tail contribution
to the gravitational-wave energy flux. To this end, we
must first consider the tail correction to the mass
quadrupole moment, as given by Eq.~(\ref{gentail}).

To leading order, we have
${{\rm I}\!\!\!{\scriptscriptstyle{{}^{-}}}}^{zz}
\propto z^2$. Because
$\dot z\approx \left(m/z\right)^{1/2}$, each time derivative has
the effect of multiplying this quantity by $m^{1/2}z^{-3/2}$.
Integration with respect to retarded time then has the effect
of multiplying the resulting expression by
by $z/\dot z\sim z^{3/2}m^{-1/2}$. As a result, we have that
${{\rm I}\!\!\!{\scriptscriptstyle{{}^{-}}}}^{zz}_{\rm Tail}
\propto\left(m/z\right)^{3/2}\,
{{\rm I}\!\!\!{\scriptscriptstyle{{}^{-}}}}^{zz}$,
which shows that the tail correction is of 1.5PN order relative
to the direct quadrupole term. Post-Newtonian corrections
to ${{\rm I}\!\!\!{\scriptscriptstyle{{}^{-}}}}^{zz}_{\rm Tail}$
will then be of 2.5PN order relative
to the same term, and can be neglected in an expansion accurate
through second post-Newtonian order. We therefore have that
in the calculation of
${{\rm I}\!\!\!{\scriptscriptstyle{{}^{-}}}}^{ij}_{\rm Tail}$,
we may put
${{\rm I}\!\!\!{\scriptscriptstyle{{}^{-}}}}^{ij}
\equiv Q^{zz}\bar{\bbox{P}}^{(2)}=
\mu z^2 \bar{\bbox{P}}^{(2)}$,
and take ${\cal C}(u)$ to represent the source's Newtonian
trajectory.

For case (A) --- infall from infinity --- the tail
term contributes to the luminosity via
\begin{equation}
{}^{(3)}{{\rm I}\!\!\!{\scriptscriptstyle{{}^{-}}}}^{ij}_{\rm Tail}
= 2m \bar{\bbox{P}}^{(2)} \int^{\infty}_0
       {}^{(5)}Q^{zz}\left(u-u'\right)
       \left[\ln\left(\frac{u'}{2{\cal S}}\right)
	 +\frac{11}{12}\right]\,du',
\label{taila}
\end{equation}
where the relation between separation $z$
and time $t$ on the Newtonian trajectory is given by
\begin{equation}
t = - \frac{4m}{3} \left( \frac{z}{2m} \right)^{3/2}.
\end{equation}
The integral in Eq.~(\ref{taila}) can be evaluated
in closed form, and the result is
\begin{equation}
{}^{(3)}{{\rm I}\!\!\!{\scriptscriptstyle{{}^{-}}}}^{ij}_{\rm Tail}
= -2\eta\left(\frac{m}{z}\right)^4\left[
\frac{71}{6}+\frac{5\pi}{\sqrt{3}}+
15\ln{\left(\frac{m}{z}\right)}
+10\ln\left(\sqrt{\frac{2}{3}}
\frac{{\cal S}}{m}\right)\right]\bar{\bbox{P}}^{(2)}.
\label{tailinf}
\end{equation}

For case (B) --- infall from a finite distance
$z_0$ --- the tail term is
\begin{equation}
{}^{(3)}{{\rm I}\!\!\!{\scriptscriptstyle{{}^{-}}}}^{ij}_{\rm Tail}
= 2m\bar{\bbox{P}}^{(2)}\int^{u(z_0)}_0
       {}^{(5)}Q^{zz}\left(u-u'\right)
       \left[\ln\left(\frac{u'}{2{\cal S}}\right)
	 +\frac{11}{12}\right]\,du'.
\end{equation}
The Newtonian trajectory is here given by
\begin{equation}
t=\frac{m}{\sqrt 2}\left(\frac{m}{z_0}\right)^{-3/2}\,g(x),
\end{equation}
where $g(x)=\sqrt{x}\sqrt{1-x}-
\arcsin{\sqrt{x}}$, with $x=z/z_0<1$.
Evaluating the integral, we obtain
\begin{equation}
{}^{(3)}{{\rm I}\!\!\!{\scriptscriptstyle{{}^{-}}}}^{ij}_{\rm Tail}
= 4\eta\left(\frac{m}{z_0}\right)^4\Biggl\{
     \left(\frac{5-4x}{x^4}\right)
     \left[\frac{11}{12}-\frac{3}{2}
     \ln{\left(\frac{m}{z_0}\right)}
-\ln\left(2\sqrt{2}\,
\frac{{\cal S}}{m}\right)\right]
+{\rm Int}(x)\Biggr\}\bar{\bbox{P}}^{(2)},
\label{tailzo}
\end{equation}
where Int$(x)$ is given by
\begin{equation}
{\rm Int}(x) = 4\int^1_x
\biggl(\frac{5-3y}{y^5}\biggr)\ln\left[g(x)-g(y)\right]\,dy.
\label{int2}
\end{equation}
In this form, Int($x$) has a very slow numerical convergence,
because the argument of the logarithm vanishes at
the lower limit. To evaluate it, we integrate by parts and
change variables, so that
\begin{eqnarray}
{\rm Int}(x) &=& -\frac{4\,(5-3x)}{x^5}\,
\sqrt{\frac{1-x}{x}}\,g(x)
\nonumber\\
& & \mbox{} + 4\int^{\pi/2}_{y}
    \frac{(24\,\sin^4w-77\,\sin^2w+55)}{\sin^{12}w}\,
    \Bigl\{\left[h(y)-h(w)\right]\,
    \ln\left[h(y)-h(w)\right]+h(w)\Bigr\}\,dw,
\end{eqnarray}
where $y=\arcsin\sqrt{x}$, and $h(w)=\sin w\cos w-w$.
This expression is much better suited for numerical calculation.

We now show that the gravitational-wave luminosity is independent
of the arbitrary parameter ${\cal S}$ appearing in
Eqs.~(\ref{tailinf}) and (\ref{tailzo}). For simplicity, we
will only consider case (A) --- infall from infinity; the
argument for case (B) is entirely analogous.

In the approach of Blanchet and Damour~\cite{blan-damour},
the energy flux at infinity is obtained by
first solving the generation of gravitational waves in the near
zone containing the source, using harmonic coordinates and
expanding about flat spacetime, and then matching this solution
outside the source to a solution of the vacuum equations in the
far zone, expressed in radiative coordinates.  In the
near-zone solution, the waves propagate along ``flat''
null cones given by
\begin{equation}
t-r=c,
\label{flat}
\end{equation}
while in the far-zone solution, they propagate along the
true null cones given by
\begin{equation}
t-r-2m \ln r=c^\prime,
\label{curved}
\end{equation}
where $c$ and $c^\prime$ are constants.   The two solutions are
matched at an intermediate distance ${\cal S}\sim \lambda
\sim r/v$, where $\lambda$ is the gravitational wavelength.
This matching is the origin of the tail term, Eq.~(\ref{tailinf}),
together with its apparent ${\cal S}$-dependence.

When evaluating the energy flux, we must trace an observation made
at time $T$ by an observer at distance $R$ back along the null cone
to the corresponding time $t_{\rm CM}$ (evaluated at $r=0$, the
center of mass) which determines the state of the source.  The
null cone that reaches the observer ($T,R$) is characterized by
$c^\prime=T-R-2m \ln R$; the match at ${\cal S}$ gives
$c^\prime=c-2m\ln{\cal S}$, and the null cone at $r=0$
is characterized by $t_{\rm CM}=c$. The result is
\begin{equation}
t_{\rm CM}=T-R-2m \ln\left(\frac{R}{\cal S}\right).
\label{shifta}
\end{equation}
We find it convenient to rescale ${\cal S}$ by $m$, and to
define retarded time $u=T-R-2m \ln(R/m)$. Then
\begin{equation}
t_{\rm CM}=u -2m \ln\left(\frac{m}{\cal S}\right).
\label{shiftb}
\end{equation}
This exhibits the explicit ${\cal S}$-dependence in the connection
between $z(t)$ and the observer. We now show that this
${\cal S}$-dependence cancels the ${\cal S}$-dependence
in the tail term.

The trajectory at Newtonian order becomes
\begin{eqnarray}
\frac{2m}{z} &=& \left(\frac{4m}{3|t_{\rm CM}|}\right)^{2/3}
 = \left\{\frac{4m}{3\left[|u|+2m\ln(m/{\cal S})\right]}\right\}^{2/3}
\simeq \left(\frac{4m}{3|u|}\right)^{2/3} \left[1-\frac{4m}{3|u|}
\ln\left(\frac{m}{{\cal S}}\right)\right] \nonumber\\
&\simeq& \frac{2m}{z'}
\left[1-\left(\frac{2m}{z'}\right)^{3/2}
\ln\left(\frac{m}{{\cal S}}\right)\right],
\label{shift}
\end{eqnarray}
where $z'=z(u)$; we have used the fact that $t_{\rm CM}<0$,
and have performed the expansions assuming $m/|u|\ll 1$.
The Newtonian contribution to
${}^{(3)}{{\rm I}\!\!\!{\scriptscriptstyle{{}^{-}}}}^{ij}$
is just the corresponding derivative of the quadrupole moment
${{\rm I}\!\!\!{\scriptscriptstyle{{}^{-}}}}^{ij}_{\rm N}
 = \mu z^2 \bar{\bbox{P}}^{(2)}$,
\begin{equation}
{}^{(3)}{{\rm I}\!\!\!{\scriptscriptstyle{{}^{-}}}}^{ij}_{\rm N} =
\frac{\eta}{2}
\left(\frac{2m}{z}\right)^{5/2}\,\bar{\bbox{P}}^{(2)},
\end{equation}
where the Newtonian expressions for $\dot z$ and $\ddot z$ were
substituted to yield a result in terms of $z$ only.
Applying the shift (\ref{shift}) to this contribution,
we obtain
\begin{eqnarray}
{}^{(3)}{{\rm I}\!\!\!{\scriptscriptstyle{{}^{-}}}}^{ij}_{\rm N}
&=& \frac{\eta}{2}
\left(\frac{2m}{z}\right)^{5/2}\,\bar{\bbox{P}}^{(2)}
\simeq \frac{\eta}{2}
\left(\frac{2m}{z'}\right)^{5/2}
\left[1-\left(\frac{2m}{z'}\right)^{3/2}
\ln\left(\frac{m}{{\cal S}}\right)\right]^{5/2}\,
\bar{\bbox{P}}^{(2)} \nonumber\\
&\simeq &
{}^{(3)}({{\rm I}\!\!\!{\scriptscriptstyle{{}^{-}}}}^{ij}_{\rm N})'
-20\,\eta\left(\frac{m}{z}\right)^4
\ln\left(\frac{m}{{\cal S}}\right)\,\bar{\bbox{P}}^{(2)}.
\label{shiftq}
\end{eqnarray}
The additional term in Eq.~(\ref{shiftq}) cancels the dependence on
$\ln(m/{\cal S})$ in the tail term, Eq.~(\ref{tailinf}).

Thus the gravitational-wave luminosity is
truly independent of the matching scale ${\cal S}$.
In practice therefore, we may assume that we have
identified the orbital variable $z$ at the correct
retarded time, and eliminate the ${\cal S}$-dependence
in the tail term by setting ${\cal S}=m$.

\subsection{Luminosity and energy loss}

We now have all the elements needed to calculate the
gravitational-wave luminosity for both types of collision, (A)
and (B). We use the acceleration (\ref{motion}) and the velocities
(\ref{veloinf}) and (\ref{velozo}) to obtain the time derivatives
of the multipole moments (\ref{multip}), including the tail terms,
Eq.~(\ref{tailinf}) and (\ref{tailzo}). We then square each term,
inserting the squares of the STF tensors, Eq.~(\ref{factors}).
Finally, we compute $\dot{E}$.

\subsubsection{Infall from infinity}

For case (A) --- infall from infinity --- the luminosity is given by
\begin{eqnarray}
\dot{E} &=&\frac{16}{15}\,\eta^2
   \left(\frac{m}{z}\right)^5
   \Biggl\{ 1-\frac{1}{7}\left(\frac{m}{z}\right)
	    \left(43-\frac{111}{2}\,\eta\right)
\nonumber\\ & & \mbox{}
   -\sqrt{2}\left(\frac{m}{z}\right)^{3/2}\left[\frac{71}{6}
       +\frac{5\pi}{\sqrt{3}}
       +15\ln{\left(\frac{m}{z}\right)}+
    5\ln\left(\frac{2}{3}\right)\right]
    -\frac{1}{3}\left(\frac{m}{z}\right)^2
     \left(\frac{1127}{9}+\frac{803}{12}\,\eta-
    112\,\eta^2\right) \Biggr\}.
\label{lumiinf}
\end{eqnarray}
In Eq.~(\ref{lumiinf}), within the curly brackets, we label the
terms of order $O(1)$, $O(m/z)$, $O[\left(m/z\right)^{3/2}]$,
and $O[\left(m/z\right)^2]$, as [N], [1PN], [1.5PN],
and [2PN], respectively.

In Fig.~1 we plot $\dot{E}/\eta^2$ as a function of $z$ for
different values of $\eta=\mu/m$, from $\eta=0$ (test-body limit)
to $\eta=0.25$ (equal-mass case). We see that $\dot{E}$ increases
monotonically with increasing $\eta$, and that the curves converge
at large $z$; for such large separations,
$\dot{E}/\eta^2 \simeq (16/15)(m/z)^5$, independent of $\eta$.
The plot also shows that $\dot{E}$ increases with decreasing $z$,
until it reaches a maximum and then starts decreasing. This behavior
signals the breakdown of the post-Newtonian expansion, as we
now discuss.

In Fig.~2a (for $\eta=0$) and 2b (for $\eta=0.25$) we
separate the luminosity~(\ref{lumiinf}) into the various
contributions corresponding to the [1PN], [1.5PN], and [2PN]
terms, and plot their ratio to the leading-order, Newtonian [N]
expression. Notice that both the [1PN] and the [2PN] contributions are
negative, while the [1.5PN] contribution is positive, because of its
logarithmic term. In both figures, for large $z$, the
post-Newtonian terms are all smaller than the Newtonian
expression. However, they are larger than might be expected
{\it a priori\/}: at $z=100m$, the [PN], [1.5PN], and [2PN]
contributions would be expected to be of order $10^{-2}$, $10^{-3}$,
and $10^{-4}$, respectively. Instead, we find that they actually
are of order $4 - 6 \times 10^{-2}$, $7 \times 10^{-2}$, and
$4 \times 10^{-3}$, respectively.
This reflects the presence of large numerical
coefficients in Eq.~(\ref{lumiinf}). As $z$ decreases, the
post-Newtonian terms increase until they become comparable to
each other and to the Newtonian expression; this occurs within
the interval $7\lesssim z/m\le10$. For $z\lesssim 7m$, the negative
post-Newtonian corrections overcome the Newtonian term, and the
luminosity formally changes sign. Evidently, the post-Newtonian
approximation is no longer reliable in this region. Finally,
comparing Fig.~2a with Fig.~2b, we see that the term which varies
the most with $\eta$ is the [1PN] contribution; the [2PN] term
varies only weakly with $\eta$, and the [1.5PN] term is independent
of $\eta$, as Eq.~(\ref{lumiinf}) directly shows.

The total energy radiated, up to time $T_f$, is given by
Eq.~(\ref{mass}),
\begin{equation}
\Delta E(z_f)=\int_{-\infty}^{T_f} \dot{E}(T)\, dT
   = -\int_{z_f}^{\infty} \dot{E}(z)\, \frac{dz}{\dot z}.
\label{massinf}
\end{equation}
The luminosity is given by~(\ref{lumiinf}) and $\dot z$ by
Eq.~(\ref{veloinf}). It is not possible to evaluate this integral
in closed form. Instead, we approximate $1/\dot z$
by its expansion in powers of $m/z$, up to the second
order, and then integrate. We obtain
\begin{eqnarray}
\Delta E(z_f)&=&\frac{16\,\sqrt{2}}{105}\,\eta^2 m\,
      \left(\frac{m}{z_f}\right)^{7/2}
   \Bigg\{1-\frac{17}{6}\left(\frac{m}{z_f}\right)
      \left(1-\frac{11}{6}\eta\right)
\nonumber\\ & & \mbox{}
   -\frac{7}{5\sqrt{2}}\left(\frac{m}{z_f}\right)^{3/2} \left[
       \frac{53}{6}+\frac{5\pi}{\sqrt{3}}+
        15\ln\left(\frac{m}{z_f}\right)
   -\ln\left(\frac{2}{3}\right) \right]
   -\frac{1}{66}\left(\frac{m}{z_f}\right)^2\left[
       \frac{81985}{36}-\frac{749}{3}\eta-
    \frac{18323}{16}\eta^2 \right]\Bigg\}.
\label{massfinal}
\end{eqnarray}

In Fig.~3 we plot $\Delta E/\eta^2 m$ as a function of
$z_f$, for different values of $\eta$. For large $z_f$, the
energy radiated behaves as its Newtonian expression,
$\Delta E/\eta^2 m \simeq (16\sqrt{2}/105)(m/z_f)^{7/2}$,
independent of $\eta$. For small $z_f$, $\Delta E/\eta^2 m$
increases with increasing $\eta$. In Fig.~3, the smallest
value of $z_f$ at which $\Delta E$ was calculated corresponds
to where the gravitational-wave luminosity formally vanishes.

\subsubsection{Infall from a finite distance}

For case (B) --- infall from a finite distance
$z_0$ --- the luminosity is given by
\begin{eqnarray} \dot{E} &=& \frac{16}{15}\,\eta^2\left(\frac{m}{z}\right)^5
    \Biggl\lgroup 1-x
    -\frac{1}{7}\left(\frac{m}{z}\right)
    \left[\left(43-\frac{111}{2}\,\eta\right)
     -x\left(116-131\,\eta\right)
     +x^2\left(71-\frac{135}{2}\,\eta\right)\right]
     \nonumber\\ & & \mbox{}
 +\sqrt{2}\left(\frac{m}{z}-\frac{m}{z_0}\right)^{1/2}
  \left\{\left(\frac{m}{z_0}\right)
     \left(x^3+4-\frac{5}{x}\right)
  \left[3\ln\left(\frac{2m}{z_0}\right) -\frac{11}{6}\right]
 +2x^3\left(\frac{m}{z_0}\right)\,{\rm Int}(x)
 \right\} \nonumber\\
& & \mbox{}
 -\frac{1}{3}\left(\frac{m}{z}\right)^2\Bigg[
  \left(\frac{1127}{9}+\frac{803}{12}\,\eta-112\,\eta^2\right)
 +\frac{1}{7}\,x
  \left(\frac{4471}{9}-\frac{15481}{3}\,\eta+2864\,\eta^2\right)
\nonumber\\
& & \mbox{}
 -\frac{1}{7}\,x^2
  \left(1870-\frac{38521}{6}\,\eta+\frac{8800}{3}\,\eta^2\right)
 +x^3 \left(83-\frac{1183}{4}\,\eta+\frac{872}{7}\,\eta^2\right)
 \Bigg]\Biggr\rgroup,
\label{lumizo}
\end{eqnarray}
where $x=z/z_0$ and Int($x$) is given by Eq.~(\ref{int2}).
This luminosity has characteristics similar to that calculated
previously. The main difference is that here, contrary to case (A),
the luminosity does not vanish at the beginning of the infall.
This, we shall now explain, is due to the fact that although the velocity
vanishes at $z=z_0$, the acceleration does not; that $\dot{E}$ does
not vanish at $z=z_0$ merely reflects the time symmetry of the
trajectory at $t=t(z_0)$.
At Newtonian order the general expression for the
energy flux, $\dot E=(8\eta^2/15)(m/r)^4(12\,v^2-11\,\dot r)$,
necessarily vanishes at a moment of stationarity ($v=\dot r=0$).
This is because it arises from an odd number (3) of
time derivatives of a mass multipole moment. At higher order,
however, contributions arising from even numbers of time
derivatives appear, e.g.,
${}^{(4)}{{\rm I}\!\!\!{\scriptscriptstyle{{}^{-}}}}^{ijk}$ in
Eq.~(\ref{lumi}). These contain terms involving the acceleration,
which does not vanish at $z_0$. As a consequence, the luminosity
also will not vanish.

As in case (A), we see here also that the luminosity changes sign
at some small value of $z$, when the negative higher-order terms
become comparable to the positive Newtonian term. This signals
the breakdown of the post-Newtonian method.

The total energy radiated, from the moment of time symmetry
to the final separation $z_f$, is given by
\begin{equation}
\Delta E(z_f) = -\int_{z_f}^{z_0} \dot{E}(z)\, \frac{dz}{\dot z},
\end{equation}
where $\dot{E}(z)$ is given by Eq.~(\ref{lumizo}) and $\dot z$
by Eq.~(\ref{velozo}). We have integrated this equation
numerically for selected values of $z_0$; the smallest value
of $z_f$ is chosen to be the one for which the luminosity
formally vanishes.

In Figs.~4a (for $\eta=0$) and~4b (for $\eta=0.25$) we plot
$\Delta E(z_f)/\eta^2 m$ as a function of the final separation
$z_f$, for different values of the initial separation $z_0$,
including the limit $z_0 = \infty$ considered previously.
We see that as $z_f$ decreases, the curves all approach
each other. This results from the fact that most of the energy is generated
near the end of the infall.

\section{Head-on collision of a small mass and
a massive black hole}

In this section we calculate the gravitational-wave
luminosity produced
during the radial infall, proceeding from rest at infinity, of
a particle with small mass into a massive, nonrotating black
hole. The smaller mass will now be denoted $\mu$, and that of the
black hole $m$. We assume $\mu \ll m$; in this limit $\mu$ and $m$ are
equivalent to the reduced mass and total mass of the previous
section.

This restriction on the mass ratio implies that the
problem considered here is a limiting case of the one
considered in the previous section, in which no
restriction was put on the masses and the nature of the
colliding objects. On the other hand, the problem considered
here can also be seen to be an extension of the one considered
in the previous section: contrary to Sec.~II, we shall here
put no restriction on the velocity of the infalling mass,
and correspondingly, put no restriction on the strength of the
gravitational field at the particle's location. While the
results of the previous section were accurate through second
post-Newtonian order, the results presented here are
accurate to all orders in $v/c$.

The stress-energy tensor associated with the infalling
particle creates a small perturbation in the gravitational
field of the nonrotating black hole (whose metric is given
by the Schwarzschild solution). Part of this perturbation
represents the Coulomb field of the infalling particle; the
remaining part represents radiative degrees of freedom, and
these propagate away from the source as gravitational waves.

We use the Teukolsky perturbation formalism \cite{Teukolsky}
to calculate $\dot{E}(T)$, the gravitational-wave luminosity
as measured near future null infinity; the luminosity is
expressed as a function of time $T$ (to be defined
precisely below). In Sec.~IV this result will be compared to
that obtained in Sec.~II using post-Newtonian theory.

The total amount $\Delta E = \int \dot{E}(T) dT$ of
gravitational-wave energy produced during the radial
infall of a particle into a Schwarzschild black hole was first
calculated by Davis, Ruffini, Press, and Price \cite{DRPP}.
This calculation was subsequently generalized to nonradial
motion by Detweiler and Szedenits
\cite{DetweilerSzedenits}.
Infall into a Kerr black hole was considered by
Sasaki and Nakamura \cite{SasakiNakamura}
(radial motion) and Kojima and Nakamura
\cite{KojimaNakamura} (nonradial motion).

The Teukolsky perturbation formalism is described
in detail in Ref.~\cite{PoissonSasaki}; we shall make
frequent use of equations contained in this paper.
In the Teukolsky formalism, the gravitational perturbations
are described by a single, complex-valued function
$\Psi_4$, which represents a particular component of the
perturbed Weyl tensor. The second-order differential
equation governing $\Psi_4$ can be completely separated
by decomposing $\Psi_4$ into Fourier modes and
spherical-harmonic components. The radial function
$R_{L M}(\omega;r)$, where $\omega$ is the angular
frequency and $L$, $M$ the spherical-harmonic indices,
satisfies an inhomogeneous, second-order, ordinary differential
equation, whose source term is constructed from the
particle's stress-energy tensor. This equation can readily
be solved by means of a Green's function.

As shown in Sec.~II of Ref.~\cite{PoissonSasaki}, the gravitational
waveforms, $h_+$ and $h_\times$, as measured by a
fictitious detector situated near
future null infinity, can be obtained from the
asymptotic behavior $R_{L M}(\omega;r\to\infty)$
of the radial function.
With a slight change in notation, Eq.~(2.14) of
Ref.~\cite{PoissonSasaki} becomes
\begin{equation}
h(T,R,\theta,\phi) = \frac{2\mu}{R} \sum_{L M}
Z_{L M}(u)\, \mbox{}_{-2}Y_{L M}(\theta, \phi).
\label{3.1}
\end{equation}
Here, $h \equiv h_+ - i h_\times$, $T$ is proper
time at the detector's location,
$R$ is the distance from the source to the
gravitational-wave detector, and
\begin{equation}
u=T-R^*=T-R-2m\ln\left(\frac{R}{2m}-1\right)
\label{3.2}
\end{equation}
is retarded time;
$T$ and $R$ are Schwarzschild coordinates, which
are distinct from the harmonic coordinates used in
the preceding section.
The functions $Z_{L M}(u)$ represent the multipole
moments of the gravitational-wave field; they are expressed
as the Fourier integrals
\begin{equation}
Z_{L M}(u) = \int \tilde{Z}_{L M}(\omega) \,
e^{-i\omega u}\, d\omega,
\label{3.3}
\end{equation}
where $\tilde{Z}_{L M}(\omega)$ will be given below. Finally,
$\mbox{}_{-2} Y_{L M} (\theta,\phi)$
are the spherical harmonics
of spin-weight $-2$ \cite{Goldbergetal}.

In terms of $h$ given above, the gravitational-wave
luminosity can be expressed as
\begin{equation}
\dot{E}(T) = \frac{R^2}{16\pi} \int
\frac{\partial \bar{h}}{\partial T}
\frac{\partial h}{\partial T}\, d\Omega,
\label{3.4}
\end{equation}
where an overbar denotes complex conjugation, and
the integration is over solid angles. Substituting
Eq.~(\ref{3.1}) into (\ref{3.4}), and using the
orthonormality of the spin-weighted spherical harmonics,
we obtain
\begin{equation}
\dot{E}(T) = \frac{\mu^2}{4\pi} \sum_{L M}
\bigl| \dot{Z}_{L M} (u) \bigr|^2,
\label{3.5}
\end{equation}
where
\begin{equation}
\dot{Z}_{L M}(u) = \int (-i\omega)\, \tilde{Z}_{L M}
(\omega)\, e^{-i\omega u}\, d\omega.
\label{3.6}
\end{equation}

It is straightforward to follow the prescription described
in Ref.~\cite{PoissonSasaki} and to calculate an expression
for $\tilde{Z}_{L M}(\omega)$ for the problem under
consideration. The first step is to describe the motion
of the particle, whose world line is taken to be a
marginally bound, radial geodesic of the Schwarzschild
spacetime. We take $\tilde{E} \equiv u_t = 1$, where
$u^\alpha$ is the particle's four-velocity, and
$\theta=0$ along the world line. The geodesic equations
are
\begin{equation}
\frac{dr}{d\tau} = - \biggl( \frac{2m}{r} \biggr)^{1/2},
\qquad
\frac{dt}{d\tau} = \frac{1}{f},
\label{3.7}
\end{equation}
where $\tau$ denotes proper time and $f=1-2m/r$.
These imply the following relation
along the world line, where $x=(r/2m)^{1/2}$:
\begin{mathletters}
\label{3.8}
\begin{equation}
t(r) = -4 m g(x)  ,
\end{equation}
\begin{equation}
g(x) =
\frac{1}{3}\, x^3 + x + \frac{1}{2}\, \ln \frac{x-1}{x+1}.
\end{equation}
\end{mathletters}

Proceeding along the lines described in Sec.~II of
Ref.~\cite{PoissonSasaki}, we obtain the following
expression for $\tilde{Z}_{L M}(\omega)$:
\begin{equation}
\tilde{Z}_{L M} (\omega) =
\tilde{Z}_{L} (\omega)\, \delta_{M0}.
\label{3.9}
\end{equation}
The fact that only the modes with $M=0$ contribute
to the radiation reflects the axial symmetry of the
problem. We also obtain
\begin{equation}
\tilde{Z}_{L}(\omega) = - \frac{i}{2m\omega}\,
\frac{\sqrt{(L-1)L(L+1)(L+2)}}{
(L-1)L(L+1)(L+2)-12im\omega}\,
\sqrt{\frac{2L+1}{4\pi}}
\frac{1}{A^{\rm in}_L(\omega)}
\int_{2m}^\infty
\frac{ \sqrt{2m/r} }{ \bigl(1+\sqrt{2m/r}\bigr)^2 }
e^{-i\omega t(r)}\, \Gamma_L (\omega)
X_L(\omega;r)\, dr.
\label{3.10}
\end{equation}
The quantities $A^{\rm in}_L(\omega)$, $\Gamma_L(\omega)$,
and $X_L(\omega;r)$ have not yet been introduced. We shall
explain their meaning in the following paragraph.

The function $X_L(\omega;r)$ is a solution to the
Regge-Wheeler equation \cite{ReggeWheeler},
\begin{equation}
\Biggl\{ \frac{d^2}{dr^{*2}} + \omega^2 +
f \biggl[ \frac{L(L+1)}{r^2} -
\frac{6m}{r^3} \biggr] \Biggr\} X_L(\omega;r) = 0,
\label{3.11}
\end{equation}
where $d/dr^* = f d/dr$. It is chosen so as to
have the asymptotic behavior
\begin{equation}
X_L(\omega;r\to 2m) \sim e^{-i \omega r^*}
\label{3.12}
\end{equation}
near the black-hole horizon. Correspondingly,
\begin{equation}
X_L(\omega;r\to\infty) \sim A^{\rm in}_L
(\omega) e^{-i\omega r^*} +
O\bigl(e^{i\omega r^*}\bigr);
\label{3.13}
\end{equation}
this equation defines the constant $A^{\rm in}_L (\omega)$
appearing in Eq.~(\ref{3.10}).
Our expression for $\tilde{Z}_L(\omega)$ also
involves the first-order differential operator
\begin{eqnarray}
\Gamma_L(\omega) &=& 2(1-3 m/r+i\omega r) r f
\frac{d}{dr} + f\left[ L(L+1)-6m/r \right]
\nonumber \\
& & \mbox{} + 2i\omega r (1-3m/r + i\omega r).
\label{3.14}
\end{eqnarray}

The strategy to calculate the gravitational-wave luminosity
is the following. For given values of $L$ and $\omega$,
the Regge-Wheeler equation (\ref{3.11}) is integrated
numerically, starting near $r=2m$ and using Eq.~(\ref{3.12})
to specify the initial conditions. The Regge-Wheeler function
and its first derivative are evaluated at values of $r$
lying in the interval $(2m,\infty)$. The
constant $A^{\rm in}_L(\omega)$ is computed using
Eq.~(\ref{3.13}). The integral to the right of
Eq.~(\ref{3.10}) is then carried out, numerically,
for the given values of $L$
and $\omega$. These steps are repeated for many relevant
values of the frequency, and the integral to the
right of Eq.~(\ref{3.6})
is evaluated for many values of $u$. Finally,
the sum over $L$ in Eq.~(\ref{3.5}) is carried out,
and $\dot{E}(u)$ is obtained for the selected values of
$u$.

The integral to the right of
Eq.~(\ref{3.10}) is actually divergent,
and must be regularized before carrying out the
program outlined in the preceding paragraph. To do this
we follow Detweiler and Szedenits \cite{DetweilerSzedenits},
who showed that the divergent behavior can be removed by
performing an integration by parts, and then discarding a
(formally infinite) boundary term. The ultimate justification
for this somewhat dangerous procedure comes from the eventual
agreement with the previous analysis of Davis
{\it et al.\/}~\cite{DRPP}, which is based on a manifestly
finite perturbation formalism.

We denote the integral to the right of Eq.~(\ref{3.10})
by $I_L(\omega)$. It is easy to see that this integral can be
split into convergent and divergent parts, so that
$I = I_{\rm conv} + I_{\rm div}$ (indices are now suppressed
for simplicity), with
\begin{equation}
I_{\rm div} = \int_{2m}^\infty
a(r)\, e^{i\omega t(r)}\, {\cal L} X(r)\, dr,
\label{3.15}
\end{equation}
where ${\cal L} = fd/dr+i\omega$, and
\begin{equation}
a(r) = \frac{2r\sqrt{r/2m}}{\bigl(\sqrt{r/2m}+1\bigr)^2}\,
\left(1-3m/r+i\omega r\right);
\label{3.16}
\end{equation}
the remainder of the integral gives $I_{\rm conv}$.

To regularize $I_{\rm div}$
we rewrite Eq.~(\ref{3.15}) as
\begin{eqnarray}
I_{\rm div} &=& \int_{2m}^\infty \biggl[
a\, e^{i\omega t}\, {\cal L} X +
\frac{d}{dr} \Bigl( b\, e^{i\omega t}\, {\cal L} X \Bigr)
\biggr]\, dr
\nonumber \\ & & \mbox{}
- b\, e^{i\omega t}\, {\cal L} X \Bigr|^\infty_{2m},
\label{3.17}
\end{eqnarray}
and seek a function $b(r)$ such that the new integral
converges. The boundary term at infinity will then be
seen to be infinite, and will be discarded; the boundary
term at the horizon will be seen to vanish. It is easy
to check that if $b(r)$ satisfies the differential
equation
\begin{equation}
\frac{db}{dr} + i\omega \biggl( \frac{dt}{dr} +
\frac{1}{f} \biggr) b + a = 0,
\label{3.18}
\end{equation}
then the new integral will converge.
A particular solution to Eq.~(\ref{3.18}) is
$b(r)=-4im\omega^{-1} (1+x)^{-1}(1+x+2im\omega x^3)$,
where $x=(r/2m)^{1/2}$. Substituting this into
Eq.~(\ref{3.17}), discarding the boundary terms, and
combining with $I_{\rm conv}$, we find that the
integral to the right of Eq.~(\ref{3.10}) can be
written in the regularized form
\begin{eqnarray}
I_L(\omega) &=& \int_{2m}^\infty
\frac{1 + im\omega\, (r/2m)^{3/2}}
{i m\omega\, (r/2m)^2}\, e^{i\omega t(r)}
\nonumber \\ & & \mbox{} \times
\left[ L(L+1) - 6m/r \right] X_L(\omega;r)\, dr.
\label{3.19}
\end{eqnarray}
It is evident that this integral is convergent at
$r=\infty$.

The numerical code written for the purpose of calculating
$\dot{E}(u)$ was built upon FORTRAN subroutines given in
{\it Numerical Recipes\/} \cite{NumRec}. The Regge-Wheeler
equation is integrated to high accuracy using the
Bulirsh-Stoer method, which is also used to evaluate
the integral (\ref{3.19}). [This integral is truncated
at a radius $r_0$ large compared with $1/\omega$;
the remainder is calculated analytically using the
asymptotic form for the Regge-Wheeler function, as
given in Eq.~(\ref{3.13}).] These steps are repeated
for many values of $\omega$; we have calculated more
than 600 points in the relevant interval
$0 \leq m\omega \leq 0.9$ \cite{DRPP}.
The Fourier transform of Eq.~(\ref{3.6})
is carried out by first selecting a value for $u$, and
then evaluating the integral using a Romberg integrator.
An interpolator is used to compute the integrand at those
values of $\omega$ which the integrator selects; these are
in general distinct from the values selected when
previously calculating $\tilde{Z}_{L}(\omega)$. These
steps are repeated for many values of $u$, lying in the
relevant interval $-80m < u < 40m$. Finally, the sum
over $L$ is carried out. Due to the rapid convergence
of the multipole expansion, only the modes with $L=2$
and $L=3$ need be included in the sum.

A plot of $\dot{E}(u)$, the gravitational-wave luminosity
as a function of retarded time, is
displayed in Fig.~5. It shows two main features:
bremsstrahlung radiation and black-hole quasi-normal ringing.
The bremsstrahlung radiation dominates the luminosity at
retarded times $u < -10m$. This radiation is generated by
the particle's infalling motion at early times (large
distances from the black hole), and propagates directly
from the particle to the detector (scattering by the
spacetime curvature is insignificant). The black-hole
ringing dominates the luminosity at retarded times
$u>-10m$. This radiation is generated by the dynamics
of the gravitational field in the vicinity of the
black-hole horizon; it is scattered many times by the
spacetime curvature, and is essentially disconnected
from the particle's motion. It is clear that quasi-normal
ringing dominates (by approximately two orders of magnitude)
the energetics of the problem.

By integrating $\dot{E}(u)$ over retarded time we were
able to reproduce the standard result for the total
energy radiated,
\begin{equation}
\Delta E = (0.0092+0.0011+\cdots) \mu^2/m,
\label{3.20}
\end{equation}
first calculated by Davis {\it et al.\/}~\cite{DRPP}.
In Eq.~(\ref{3.20}), the first term comes from the
$L=2$ mode, while the second corresponds to $L=3$.
The total coefficient, as computed in Ref.~\cite{DRPP}
and reproduced here, is equal to 0.0104.

\section{Comparison between post-Newtonian and
perturbation-theory results}

The purpose of this section is to compare the results
obtained in Sec.~II using post-Newtonian theory to those
obtained in Sec.~III using black-hole perturbation
theory. More precisely, we shall consider the
$\eta\to 0$ limit (where $\eta=\mu/m$) of Eq.~(\ref{lumiinf})
and compare it to the numerical results displayed in
Fig.~5. While the numerical results are valid to
all orders in the post-Newtonian expansion, the
analytic expression is accurate only through second
post-Newtonian order.

In the limit of small mass ratios, Eq.~(\ref{lumiinf})
takes the form
\begin{equation}
\dot{E}(T_h) =\frac{16}{15} \eta^2
\biggl( \frac{m}{r_h} \biggr)^5 \Biggl\{
1 - \frac{43}{7} \biggl( \frac{m}{r_h} \biggr)
- \sqrt{2} \biggl[ \frac{71}{6} + \frac{5\pi}{\sqrt{3}}
+ 15 \ln\biggl(\frac{m}{r_h}\biggr)
+5 \ln\biggl( \frac{2}{3} \biggr) \biggr]
\biggl(\frac{m}{r_h}\biggr)^{3/2} -
\frac{1127}{27} \biggl( \frac{m}{r_h} \biggr)^2 \Biggr\}.
\label{4.1}
\end{equation}
For convenience, we have made slight changes
to the notation used in Sec.~II: we have
replaced $z$ by $r_h$, where the subscript indicates
that the coordinates are harmonic. The gravitational-wave
luminosity is expressed as a function of time $T_h$,
which operationally has the same meaning as in Sec.~III:
it corresponds to proper time as measured by a static
gravitational-wave detector situated near future null
infinity.

Equation (\ref{4.1}) is incomplete without the relationship
between $T_h$ (observer harmonic time at large distances),
Which appears to the left, and $r_h$, which
appears to the right. This relationship is discussed in
Sec.~IIC: If $r_h(t_h)$ describes the particle's world line
in harmonic coordinates, then the quantity $r_h$ appearing
to the right of Eq.~(\ref{4.1}) is $r_h(u)$, where $u$ is
the retarded time
\begin{equation}
u = T_h - R_h - 2m\ln \biggl( \frac{R_h}{m} \biggr).
\label{4.2}
\end{equation}
Here, $R_h \gg m$ is the distance from the
detector to the system's center
of mass (which here is identical to the fixed position of
the large mass $m$). Notice that Eq.~(\ref{4.2}) takes the
same form as in Eq.~(\ref{3.2}), in the limit of large $R/m$,
apart from two slight
differences. The first is that Eq.~(\ref{3.2}) involves
Schwarzschild coordinates, $T_S$ and $R_S$, instead of
harmonic coordinates. The relationship between these
coordinates is
\begin{equation}
T_S = T_h + \mbox{const.}, \qquad R_S = R_h + m.
\label{4.3}
\end{equation}
The second is that in Eq.~(\ref{4.2}) $R_h$ is divided
by $m$, while in Eq.~(\ref{3.2}) $R_S$ is divided by $2m$.
This difference originates in the freedom in selecting the
value of the parameter $\cal S$, as was discussed in
Sec.~IIC. This corresponds to the freedom in choosing the
origin of the retarded time $u$. In particular, the origin
of time might be chosen differently in the post-Newtonian and
perturbation-theory calculations.

Because of this ambiguity in the origin of $u$, it would
be unwise to compare directly Eq.~(\ref{4.1}) with
Fig.~5. A much better way of carrying out the comparison
is to re-express the luminosity function in terms
of a fully unambiguous parameter. We
shall choose to parametrize the world line with the
Schwarzschild coordinate $r_S$, and express the luminosity
in terms of this parameter. It might appear that this
re-expression has already been effected (apart from
the transformation $r_S = r_h + m$) in Eq.~(\ref{3.2}).
We shall see, however, that the right-hand side of
Eq.~(\ref{3.2}) is {\it not\/} the desired expression.

Let us begin with the perturbation-theory results for
$\dot{E}(u)$, and let us adopt a mapping between $u$,
the time coordinate running along future null infinity, and
$r_S$, the adopted parameter along the particle's
world line. The simplest and most natural mapping is
depicted in Fig.~6: we connect the event labeled by $r_S$
on the world line and the event labeled by $u$
on future null infinity with a radial, outgoing null
geodesic. This null geodesic is taken to propagate in the
direction directly opposite to the direction in which the
particle moves. Radial, outgoing null geodesics are such that
$t_S - r_S - 2m\ln(r_S/2m-1)$ is constant along them.
The relation $u(r_S)$ therefore follows directly from
Eq.~(\ref{3.2}), with $r_S$ substituted for $R$, and
$t_S(r_S)$, given by Eq.~(\ref{3.8}), substituted for $T$.
The result is
\begin{equation}
u(r_S) = -4m\left[
\frac{1}{3} x^3 + \frac{1}{2} x^2 + x + \ln (x-1) \right],
\label{4.4}
\end{equation}
where $x=(r_S/2m)^{1/2}$. Using Eq.~(\ref{4.4}), the
luminosity function can be plotted as a function of
$r_S$, the parameter along the world line. This plot
is displayed as a solid curve in Fig.~7.

The figure shows that the numerical curve tends to
become pointwise unreliable as the radius increases
beyond $r_S = 20m$: the computed curve undergoes small
oscillations about a mean curve which presumably
represents the true luminosity function. These
oscillations are due to numerical error. More precisely,
they are a consequence of the fact that the integration
over angular frequencies, cf.~Eq.~(\ref{3.6}), must
necessarily be cut off at some finite upper bound $\omega_0$.
Reducing this upper bound produces larger oscillations
about approximately the same mean curve.

The mapping between world line and future null infinity
adopted in Eq.~(\ref{4.1}) is different from the one
constructed in the preceding paragraph. This is understood
from Eq.~(\ref{4.2}), which involves the distance $R_h$ to
the system's center of mass, instead of the distance
to the particle itself. Therefore our plot for
$\dot{E}(r_S)$ cannot be compared directly with
Eq.~(\ref{4.1}). To carry out a meaningful comparison,
we must first map $r_A$, the position of the particle at the retarded
time of the center of mass,
into the true-retarded position $r_B$, as illustrated
in Fig.~8. (The coordinates used here are the harmonic
coordinates, but we suppress our use of the subscript $h$.)
We now turn to this task.

Figure~8 illustrates how to construct the mapping from
$r_A$ to $r_B$. As mentioned previously, the coordinates
used for this purpose are the harmonic coordinates. Let
$t=p(r)$ represent the particle's world line, and let
$t=q(r)+u$ represent a light ray propagating in the opposite
direction and meeting future null infinity at retarded time
$u$. The center of mass is assumed to have the fixed position
$r=0$. The radius $r_A$ is determined by inverting the equation
\begin{equation}
p(r_A) = q(0) + u.
\label{4.5}
\end{equation}
The radius $r_B$, on the other hand, is obtained by solving
\begin{equation}
p(r_B) = q(r_B) + u.
\label{4.6}
\end{equation}
By eliminating $u$ from these equations, we obtain the
desired (implicit) relationship between $r_A$ and $r_B$:
\begin{equation}
p(r_A) = p(r_B) - q(r_B) + q(0).
\label{4.7}
\end{equation}
Our goal now is to transform Eq.~(\ref{4.7}) into
something more concrete.

We need to derive appropriate expressions for $p(r)$
and $q(r)$. The situation for $q(r)$ is quite simple.
An expression for it can be derived from
Eqs.~(\ref{flat})--(\ref{shiftb}). With $\cal S$
consistently set equal to $m$, we obtain simply
\begin{equation}
q(r) = r.
\label{4.8}
\end{equation}

An expression for $p(r)$ can be derived by integrating
$dr/dt$, which was written down in Sec.~IIB. In the limit
$\eta \to 0$, Eq.~(\ref{veloinf}) becomes
\begin{equation}
\frac{dr}{dt} = - y^{1/2} \left( 1 - \frac{5}{4} y +
\frac{27}{32} y^2 \right),
\label{4.9}
\end{equation}
where $y=2m/r$; Eq.~(\ref{4.9}) is valid through
second post-Newtonian order. Integrating
Eq.~(\ref{4.9}) would give an expression for
$p(r)$ also accurate to second post-Newtonian order.
This would imply that the mapping is carried out
with the same degree of accuracy as the calculation
of $\dot{E}$, as given in Eq.~(\ref{4.1}). However,
we have here the opportunity of using an
expression for $dr/dt$ that is more accurate
than Eq.~(\ref{4.9}). This comes about because
in the $\eta \to 0$ limit, the equations of motion
are given exactly by the Schwarzschild expressions
(\ref{3.7}). Once written in terms of the harmonic
coordinates, these equations imply
\begin{equation}
\frac{dr}{dt} = -y^{1/2} \left(1-
{\textstyle \frac{1}{2}} y \right)
\left(1+
{\textstyle \frac{1}{2}} y \right)^{-3/2}.
\label{4.10}
\end{equation}
Expanding this in powers of $y$ \cite{foot1}, we obtain
\begin{eqnarray}
\frac{dr}{dt} &=& -y^{1/2} \biggl( 1 - \frac{5}{4} y
+ \frac{27}{32} y^2 - \frac{65}{128} y^3
\nonumber \\ & & \mbox{}
+ \frac{595}{2048} y^4 - \frac{1323}{8192} y^5 + \cdots \biggr),
\label{4.11}
\end{eqnarray}
which generalizes Eq.~(\ref{4.9}). For the purpose
of carrying out the mapping we will deal
with Eq.~(\ref{4.11}), whose integration gives $p(r)$:
\begin{eqnarray}
p(r) &=& -\frac{4m}{3}\, y^{-3/2} \biggl( 1 + \frac{15}{4} y
- \frac{69}{32} y^2 - \frac{45}{128} y^3
\nonumber \\ & & \mbox{}
- \frac{1089}{10240} y^4 - \frac{2169}{57344} y^5 + \cdots \biggr).
\label{4.12}
\end{eqnarray}

The mapping between $r_A$ and $r_B$ is given by
Eqs.~(\ref{4.7}) and (\ref{4.8}): $p(r_A) = p(r_B) - r_B$.
This equation could be inverted numerically to yield $r_A(r_B)$.
However, the following approximate, analytic inversion proves to
be sufficiently accurate for our purposes. First, we define an
auxiliary variable $Z \equiv [-3p(r_A)/4m]^{-2/3}$. Next,
Eq.~(\ref{4.12}) is used to express $Z$ as a power
series in $y_A \equiv 2m/r_A$; this series is then inverted
to yield $y_A$ as a power series in $Z$:
\begin{equation}
y_A = Z \Bigl( 1 + \frac{5}{2} Z + \frac{13}{4} Z^2
   - \frac{65}{24} Z^3 - \frac{5821}{240} Z^4 + \cdots \Bigr).
\label{4.13}
\end{equation}
Finally, we use our mapping equation to write
$Z = [-3p(r_B)/4m + 3r_B/4m]^{-2/3}$; substitution of
Eq.~(\ref{4.12}) then gives $Z$ in terms of
$y_B \equiv 2m/r_B$:
\begin{eqnarray}
Z &=& y_B \biggl( 1 + \frac{3}{2} {y_B}^{1/2} + \frac{15}{4} y_B
   - \frac{69}{32} {y_B}^2 - \frac{45}{128} {y_B}^3
\nonumber \\ & & \mbox{}
   - \frac{1089}{10240} {y_B}^4 - \frac{2169}{57344} {y_B}^5
   + \cdots \biggr)^{-2/3}.
\label{4.14}
\end{eqnarray}

We have obtained the desired inversion: a value of $r_B$ is
selected, $Z$ is calculated using Eq.~(\ref{4.14}), and
$r_A$ is then obtained from Eq.~(\ref{4.13}). It is
this radius which must be substituted into Eq.~(\ref{4.1})
in order to obtain the gravitational-wave luminosity. This
can then be plotted against the Schwarzschild radius
$r_S = r_B + m$, which gives the dotted curve in Fig.~7.

The accuracy of the post-Newtonian expression can be ascertained
from the figure. We find that the post-Newtonian value differs
from the numerical one by a factor of approximately 2.5
at $r_S = 10m$, 1.5 at $r_S = 20m$, and 1.3 at $r_S = 30m$.
We see that the post-Newtonian values tend to converge
at large distances to the exact values, but that the rate
of convergence is slow. We will explore this issue of the
convergence of the post-Newtonian values more fully in the
following section.

\section{Accuracy of the post-Newtonian expansion}

We have seen in the preceding section that the
post-Newtonian expression for the gravitational-wave
luminosity, as given by Eq.~(\ref{4.1}), only
approximately reproduces the exact, numerical curve,
as represented in Fig.~7. And indeed, the degree of
accuracy is far less than could be expected:
An expression valid
to second post-Newtonian order could be expected to
have a fractional accuracy of order $(m/r)^3$; for
$r=10m$ this is of order $10^{-3}$. Instead, comparison
with the numerical results shows that the fractional accuracy
is actually of order $0.5$.

In the previous section, the comparison between the
post-Newtonian and perturbation-theory results could
be carried out for $r_S < 30m$ only. The purpose of
this section is to push the comparison to larger
values of the radius, in spite of the fact that the
numerical results are not reliable beyond $r_S = 30m$.

To this end, we consider the following calculation,
based on the hybrid formalism proposed by Kidder, Will,
and Wiseman \cite{kidder}. We calculate the
gravitational-wave luminosity associated with the radial
infall of a particle with small mass into a much more massive,
spherically symmetric object (not necessarily a black hole).
The masses are denoted $\mu$ and $m$ respectively, and
we assume $\mu \ll m$. We do so by using the multipole expansion
of Eq.~(\ref{lumi}), and by employing the {\it exact\/}, Schwarzschild
equations of motion given by Eq.~(\ref{4.10}). Taking
advantage of the rapid convergence of the multipole expansion,
only the mass quadrupole term will be considered;
tail effects will also be ignored.

The calculation proceeds as follows.
In a Cartesian coordinate system based on the harmonic
coordinates, the system's trace-free quadrupole moment
can be expressed as
${{\rm I}\!\!\!{\scriptscriptstyle{{}^{-}}}}^{ij} =
\mu r^2 \bar{\bbox{P}}^{(2)}$,
where $\bar{\bbox{P}}^{(2)} = \mbox{diag}(-1/3,-1/3,2/3)$.
Taking three time derivatives, we obtain
\begin{equation}
\mbox{}^{(3)} {{\rm I}\!\!\!{\scriptscriptstyle{{}^{-}}}}^{ij}
= 2\mu \biggl( r\, \frac{d^3r}{dt^3} + 3
\frac{dr}{dt}\, \frac{d^2r}{dt^2} \biggr)
\bar{\bbox{P}}^{(2)}.
\label{5.1}
\end{equation}
Use of Eq.~(\ref{4.10}) and substitution
into the first term of Eq.~(\ref{lumi}) finally yields
\begin{equation}
\dot{E} = \frac{16}{15}\, \eta^2
\biggl( \frac{m}{r} \biggr)^5 {\cal R}^2,
\label{5.2}
\end{equation}
where
\begin{eqnarray}
{\cal R} &=& \left(1-\frac{m}{r}\right)
\left(1+\frac{m}{r}\right)^{-13/2}
\nonumber \\ & & \times
\left[1+5\left(\frac{m}{r}\right)
-29\left(\frac{m}{r}\right)^2
+15\left(\frac{m}{r}\right)^3\right].
\label{5.3}
\end{eqnarray}
Equation (\ref{5.2}) with ${\cal R} = 1$ gives the standard
quadrupole-formula result; the factor ${\cal R}^2$ is the
relativistic correction. Notice that ${\cal R}$ tends
toward unity at large $r$, and that ${\cal R}=0$ at
$r=m$ (the event horizon) and $r=3m$.

Equation (\ref{5.2}) is to be interpreted as the fully
relativistic analogue of Eq.~(\ref{4.1}). This
expression can be compared with the exact,
numerical curve of Fig.~7, provided that the
mapping illustrated in Fig.~8 is carefully carried out.
The result is displayed as the dotted curve in
Fig.~9. We see that Eq.~(\ref{5.2}) reproduces
the numerical results to very high accuracy for
$r_S > 7m$. In particular, this expression does
much better than its post-Newtonian analogue.
This agreement should not be considered to be of
deep physical significance; it is essentially a
fluke. However, it will allow us to use Eq.~(\ref{5.2})
as a tool for determining the large-distance accuracy
of the post-Newtonian results, as we now explain.

Although the numerical curve cannot, because of numerical
error, be reliably constructed for radii larger than
approximately $30m$, it is most plausible that
the agreement with Eq.~(\ref{5.2}) would persist
at larger values of $r$. It therefore appears appropriate
to investigate the large-distance convergence property of
the post-Newtonian luminosity, with respect to the exact
curve, by comparing Eq.~(\ref{4.1}) directly with
Eq.~(\ref{5.2}). This comparison, which has the major
advantage of not involving the mapping illustrated in
Fig.~8, reveals that the relative difference between
the post-Newtonian values and the exact ones is
20\% at $r_h \simeq 45m$, 10\% at $r_h \simeq 75m$,
and 1\% at $r_h \simeq 400m$. These results establish
the slow convergence of the post-Newtonian expansion.

\acknowledgments
Part of this work was carried out with the aid of the symbolic
manipulator MACSYMA. This research was supported in part
by the National Science Foundation under Grant
No.~PHY 92-22902 and the National Aeronautics and Space Administration
under Grand No. ~NAGW 3874.

\appendix
\section*{Cancellation of post-Newtonian terms}

The purpose of this Appendix is to point out a remarkable
cancellation of terms occurring in the calculation of the
gravitational-wave luminosity in the case of a head-on
collision proceeding from rest at infinity.

Substituting Eq.~(\ref{veloinf}) into Eqs.~(\ref{multip})
we find that the mass multipole moments can be expressed
as expansions in powers of $m/z$ only. To second
post-Newtonian order, the generic form is
\begin{mathletters}
\label{coeffs}
\begin{eqnarray}
{{\rm I}\!\!\!{\scriptscriptstyle{{}^{-}}}}^{zz}
  & = & \mu z^2\left[A_{2,0}+A_{2,1}\left(\frac{m}{z}\right)
	+A_{2,2}\left(\frac{m}{z}\right)^2\right], \\
{{\rm I}\!\!\!{\scriptscriptstyle{{}^{-}}}}^{zzz}
& = & \mu z^3\left[A_{3,0}+A_{3,1}
\left(\frac{m}{z}\right)\right],\\
{{\rm I}\!\!\!{\scriptscriptstyle{{}^{-}}}}^{zzzz}
& = & \mu z^4\,A_{4,0}.
\end{eqnarray}
\end{mathletters}
Assuming that such an expansion holds at least to order
$(m/z)^n$, we write the mass $n$-pole moment
in the form
\begin{eqnarray}
{{\rm I}\!\!\!{\scriptscriptstyle{{}^{-}}}}^{zz\cdots}
&=& \mu z^n \Biggl\{A_{n,0}+A_{n,1}\left(\frac{m}{z}\right)+
\cdots
\nonumber \\ & & \mbox{}
+A_{n,n}\left(\frac{m}{z}\right)^n
+O\biggl[\left(\frac{m}{z}\right)^{n+1}\biggr]\Biggr\}.
\label{genmulti}
\end{eqnarray}
We have represented the first few moments in Fig.~10,
in which a row corresponds to a given multipole order, and
a column to a given power of $m/z$ in the expansion.
The thick dashed line separates the coefficients needed in
order to express the luminosity accurately through 2PN order;
these appear to the left of the boundary.

First, we observe that in Eq.~(\ref{genmulti}), the term of order
$O[\left(m/z\right)^n]$ is just the constant $A_{n,n}\,\mu\,m^n$.
The contribution from this term to the energy flux therefore
vanishes when time derivatives are taken. We designate
this cancellation in Fig.~10 by the darkly shaded diagonal
beginning at $A_{2,2}$.

Second, because $\dot z\propto z^{-1/2}$ to leading order, the
$k^{\rm th}$ time derivative of a term of the form
$A_{l,n}\,z^{l-n}$, with $l>n$, goes like $z^{(l-n-3k/2)}$.
Consequently, if $l-n=0_{\rm mod~3}$, this term becomes a
constant after $k_0=2(l-n)/3$ derivatives, and will vanish after
an additional derivative is taken. But since each $l$-pole
moment is differentiated $l+1$ times to calculate the energy
flux, this term will vanish whenever $k_0\le l-1$, or
$l+2n\ge 3$. Such terms are indicated in Fig.~10 by
the lightly shaded diagonals.

This does not imply, however, that contributions from the PN
order corresponding to a shaded region do not appear at all
in the time-derivatives of the corresponding STF moment. If
the term in question is a higher-order correction to the moment
(e.g., $A_{2,2}$ or $A_{4,1}$), then contributions to
${}^{(3)}{{\rm I}\!\!\!{\scriptscriptstyle{{}^{-}}}}^{ij}$
of the same order could be generated by correction terms in the
equations of motion, applied to derivatives of lower-order terms.
Only if the term is a leading-order term (e.g., $A_{3,0}$ or
$A_{6,0}$) is the contribution at that order identically zero.
It is interesting to note that because of this effect, the
coefficient $A_{2,2}$ is not explicitly needed in calculating
the energy flux to 2PN order. This coefficient has only recently
been calculated~\cite{bdiww}.

In the case of infall from a finite distance, the argument
used above fails. To leading order, the velocity is now
\begin{equation}
\dot z\propto \left( \frac{m}{z} -
\frac{m}{z_0} \right)^{1/2},
\end{equation}
and the 2PN expansion for
${{\rm I}\!\!\!{\scriptscriptstyle{{}^{-}}}}^{ij}$
contains powers of both
$\left(m/z\right)$ and $\left(m/z_0\right)$. For example, terms of the
form $m^2/z z_0$ and $\left(m/z_0\right)^2$ appear in the 2PN
expansion for the quadrupole moment; when multiplied by $\mu z^2$,
these terms are not killed by time derivatives.


\begin{figure}
\caption{Gravitational-wave luminosity, in units of $\eta^2$,
as a function of separation $z$, for case (A): infall from
infinite initial separation. We show the luminosity curves
for $\eta= \{0,0.15,0.25\}$.}
\end{figure}

\begin{figure}
\caption{A plot of the absolute value of the 1PN, 1.5PN, and 2PN
contributions to the luminosity divided by the Newtonian
expression. The 1PN and 2PN contributions are negative; the
1.5PN contribution is positive. Fig.~2a shows the test-body
limit ($\eta=0$), and Fig.~2b shows the equal-mass case
($\eta=0.25$).}
\end{figure}

\begin{figure}
\caption{Total energy radiated during an
infall from infinite initial separation to final separation
$z_f$. We show curves for $\eta=\{0,0.15,0.25\}$.}
\end{figure}

\begin{figure}
\caption{Total energy radiated during an
infall from initial separation $z_0$ to final separation
$z_f$, including the limit $z_0 = \infty$. We show curves for
$z_0/m=\{15, 30, 50, \infty \}$.  Fig.~4a
shows the test-body limit ($\eta=0$); Fig.~4b shows the
equal-mass case ($\eta=0.25$).}
\end{figure}

\begin{figure}
\caption{A plot of $\dot{E}/\eta^2$ (on a logarithmic scale)
as a function of retarded time $u/m$, as calculated using
perturbation theory. The luminosity represents bremsstrahlung
radiation for $u/m < -10$ and black-hole quasi-normal ringing
for $u/m > -10$.}
\end{figure}

\begin{figure}
\caption{Conformal diagram representing the Schwarzschild
spacetime. Shown are: past future null infinity (lower
diagonal with positive slope), future null infinity (higher
diagonal with negative slope), the future horizon (higher
diagonal with positive slope), the past horizon (lower
diagonal with negative slope), and the singularity
(broken horizontal). We also represent the world line
of the infalling particle, parametrized by the Schwarzschild
coordinate $r_S$, and the radial, outgoing null geodesic
reaching future null infinity at retarded time $u$.}
\end{figure}

\begin{figure}
\caption{A plot of $\dot{E}/\eta^2$ (on a logarithmic scale)
as a function of $r_S/m$, the parameter along the particle's
world line. The solid curve represents the exact, numerical
results obtained using perturbation theory. The dotted curve
represents the post-Newtonian approximation given by Eq.~(4.1).}
\end{figure}

\begin{figure}
\caption{The mapping between $r_A$, the center-of-mass-retarded
position, and $r_B$, the true-retarded position. The curve
$t=p(r)$ represents the world line of the particle. The
curve $t=q(r)+u$ represents an outgoing null ray propagating
in the opposite direction, reaching future null infinity at
retarded time $u$. The center of mass is located
at the center $r=0$.}
\end{figure}

\begin{figure}
\caption{A plot of $\dot{E}/\eta^2$ (on a logarithmic scale)
as a function of $r_S/m$, the parameter along the particle's
world line. The solid curve represents the exact, numerical
results obtained using perturbation theory. The dotted curve
represents the quadrupole-relativistic approximation given
by Eq.~(5.2).}
\end{figure}

\begin{figure}
\caption{Cancellations of terms in the case of infall from
infinity. Each row represents a multipole order, with
a common factor $\mu\,z^i$, while each column labels the
contribution of order $O(m/z)^j$ to the multipole moment
in a post-Newtonian expansion.  Tail terms are ignored.
A consistent expansion of
the luminosity up to a certain order $n$ involves the
coefficients included in the upper-left corner of the table
(the requirement for an expansion of the luminosity accurate
to 2PN order is marked by the heavy dashed line). The
coefficients which have currently been calculated explicitly are
indicated in bold face.}
\end{figure}


\begin{references}
\bibitem{foot1} This equation could be integrated exactly,
and the result would be equivalent to Eq.~(3.8). However,
the exact result implies that the system's center of mass
lies within the black-hole horizon, which means that the
mapping depicted in Fig.~4.3 is not even defined in this
exact limit. It is therefore desirable to expand Eq.~(4.10)
in powers of $y$, and take advantage of the fact that this
can be done to quite a high degree of accuracy for $r \gg m$.
\end{references}
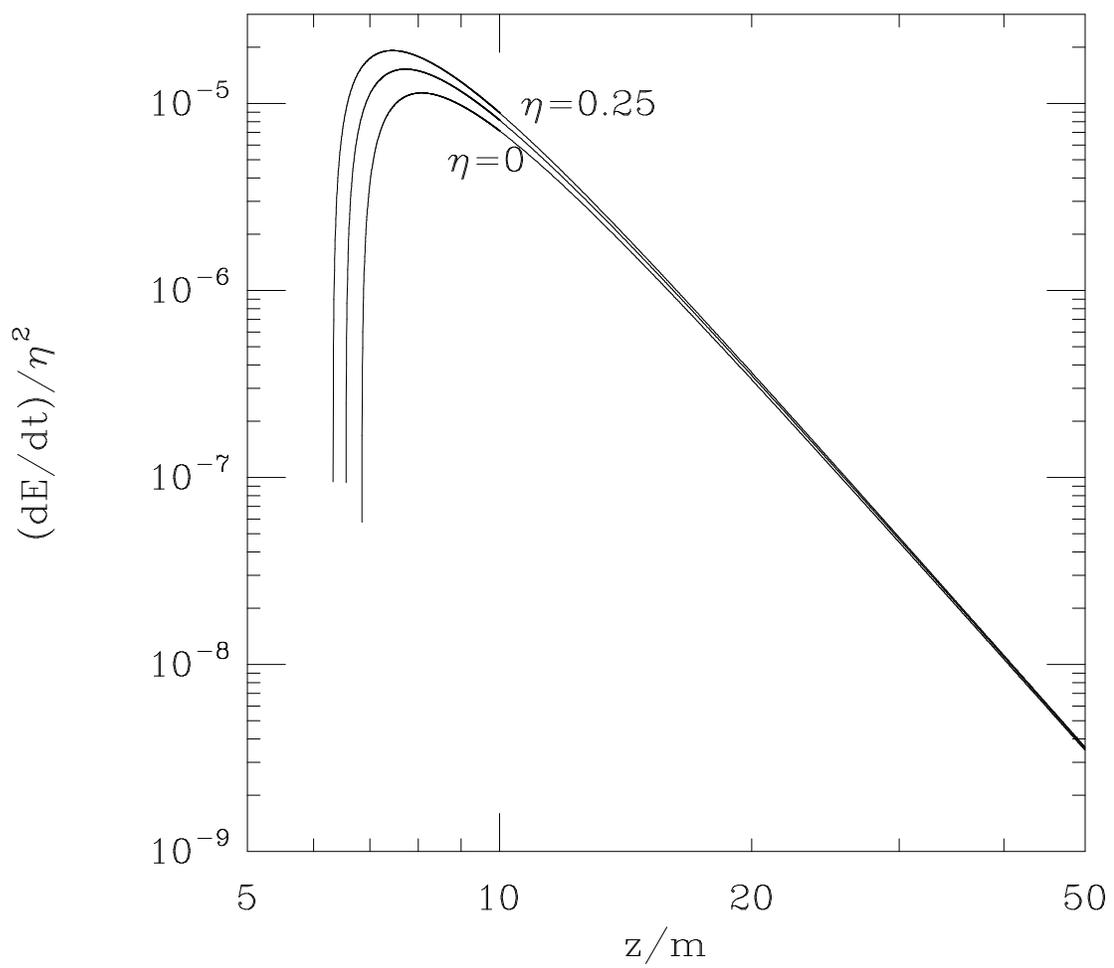
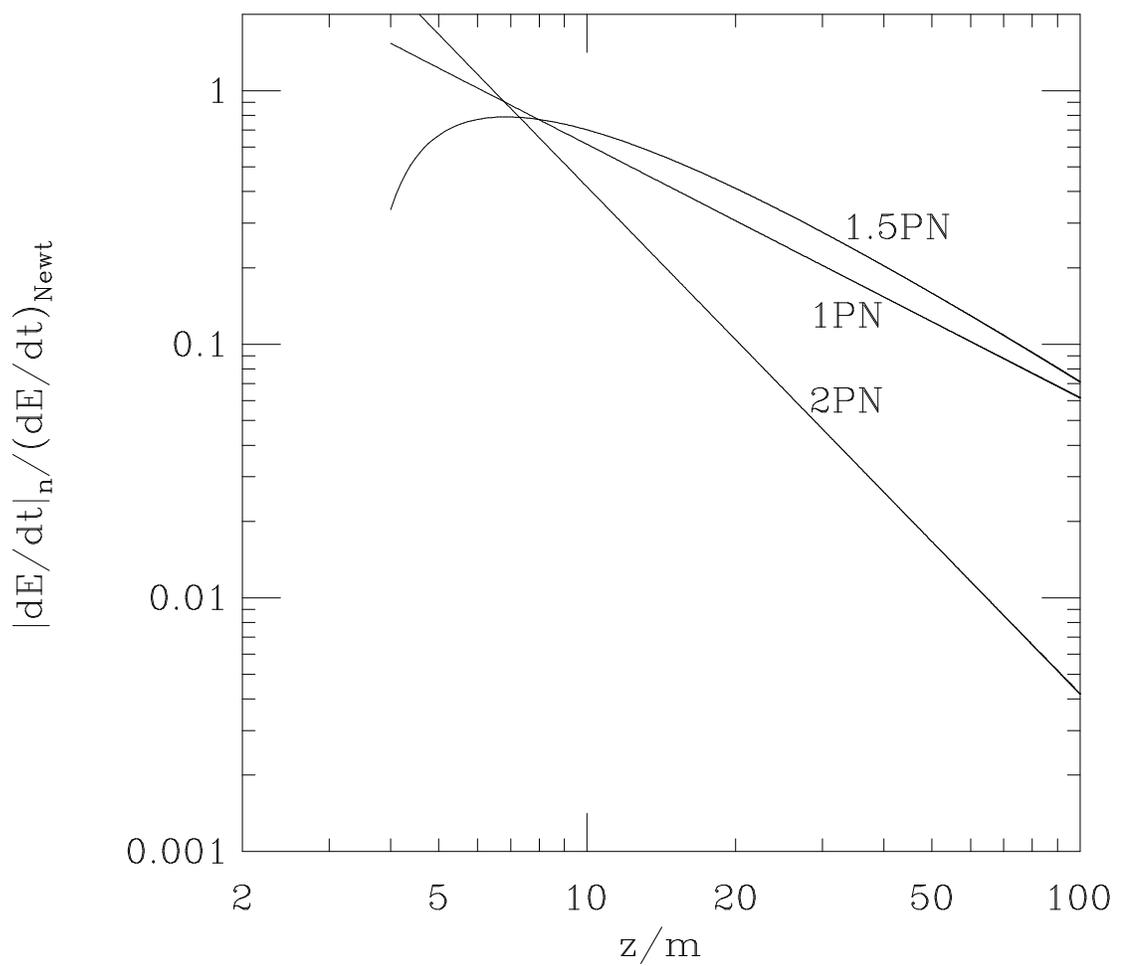
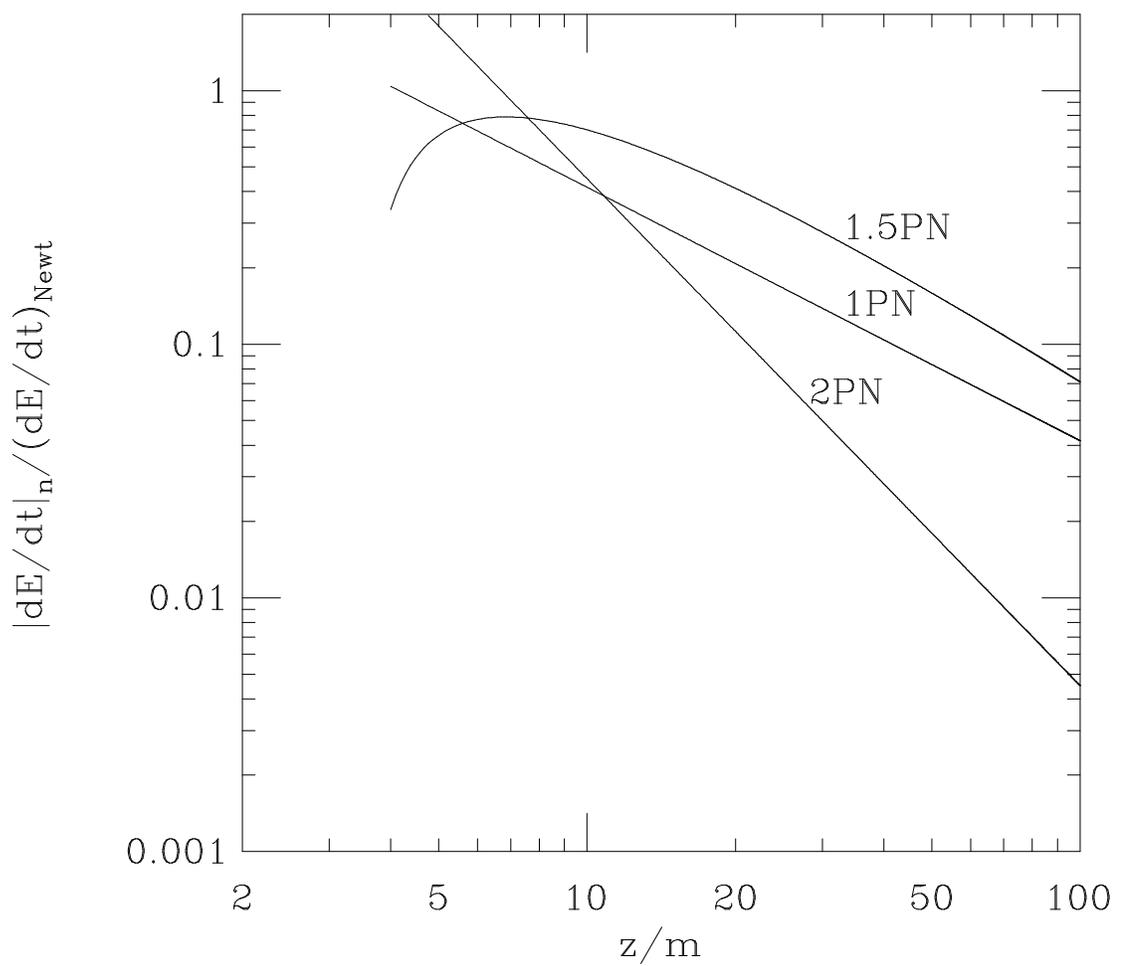
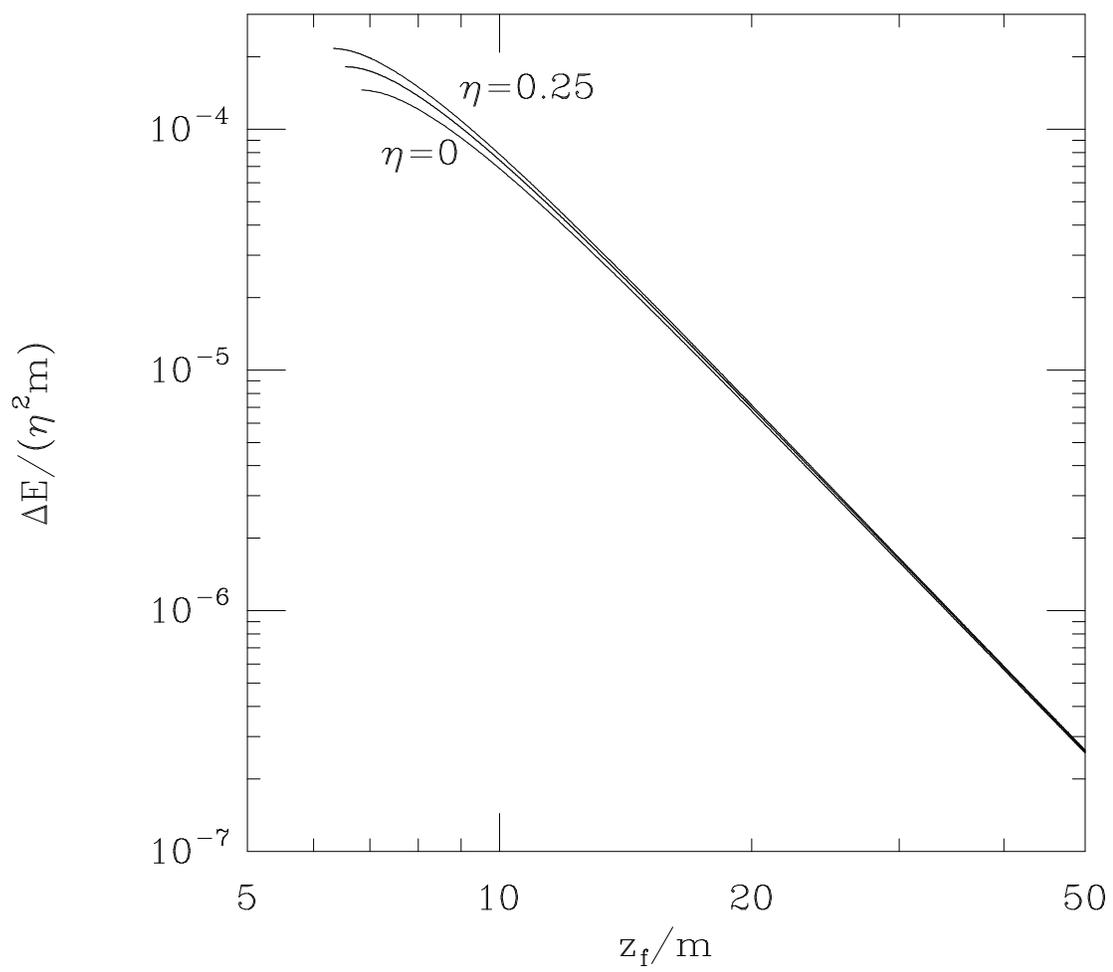
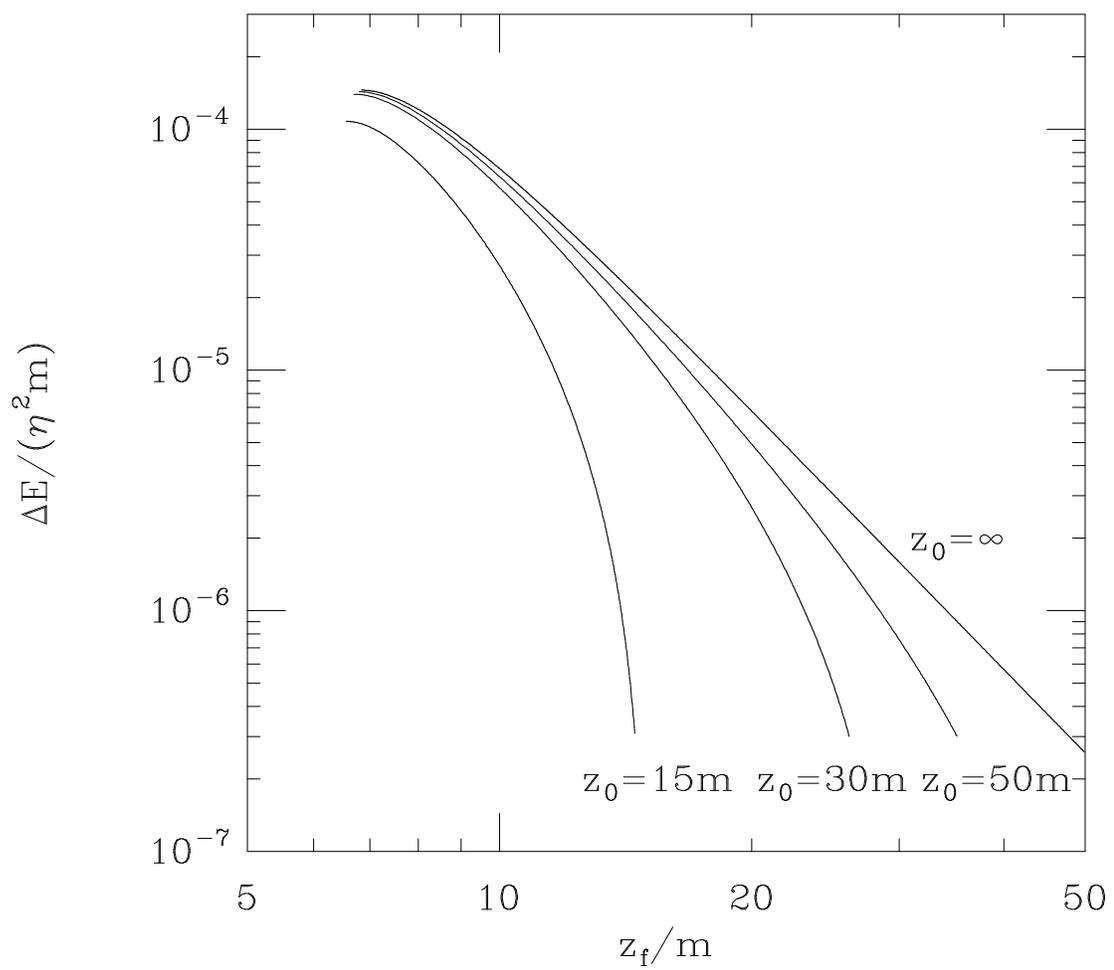
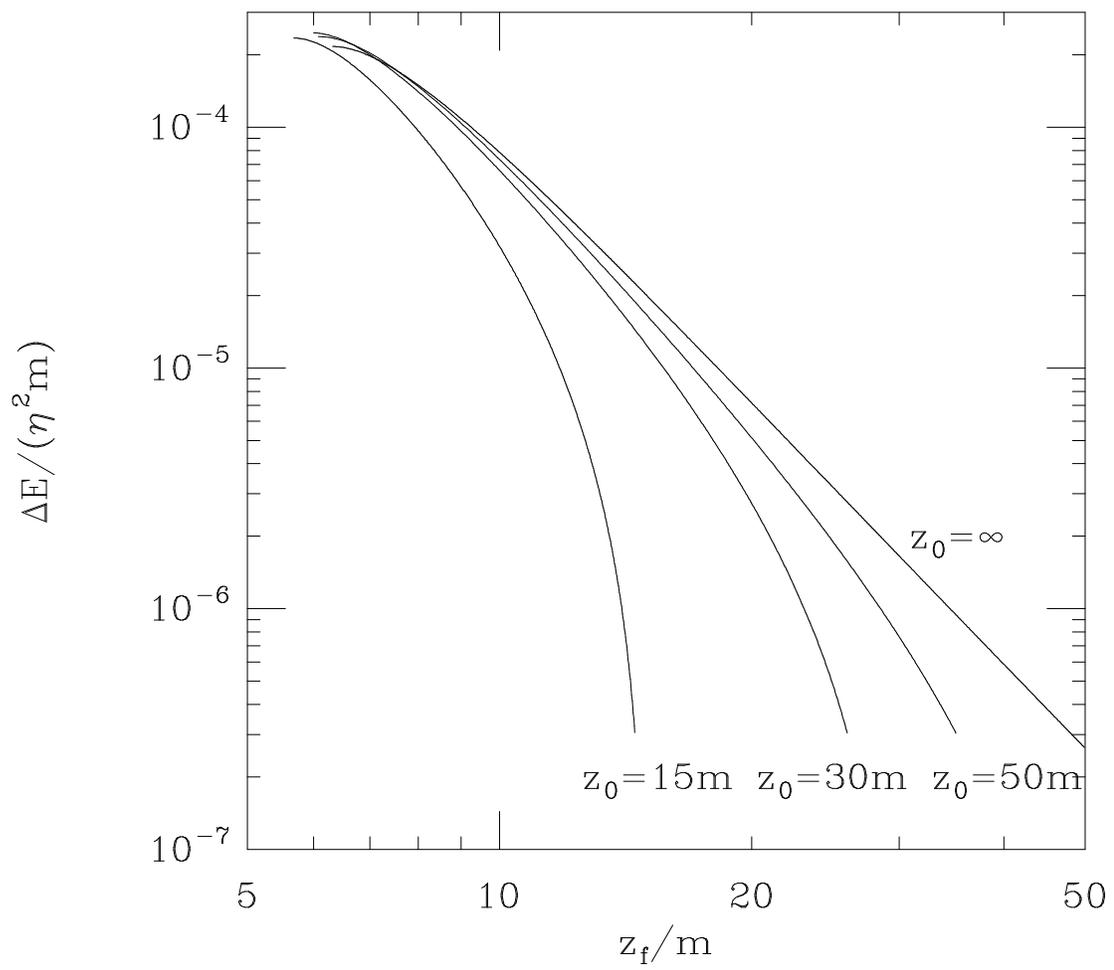
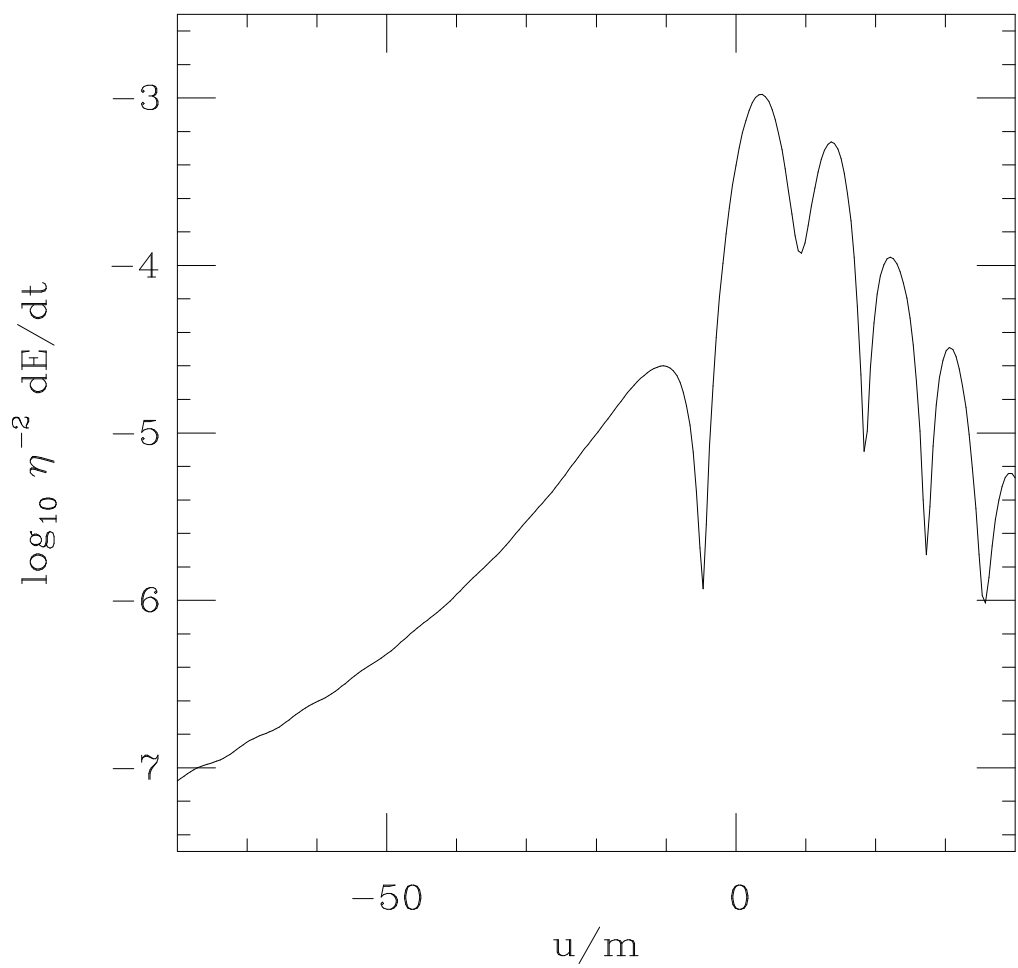
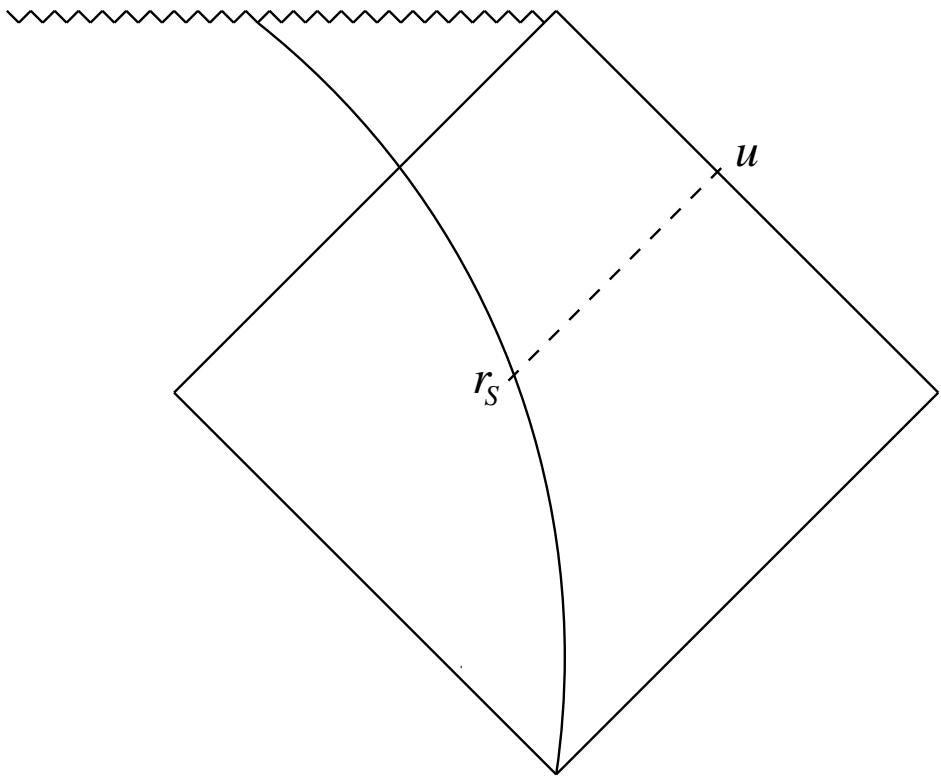
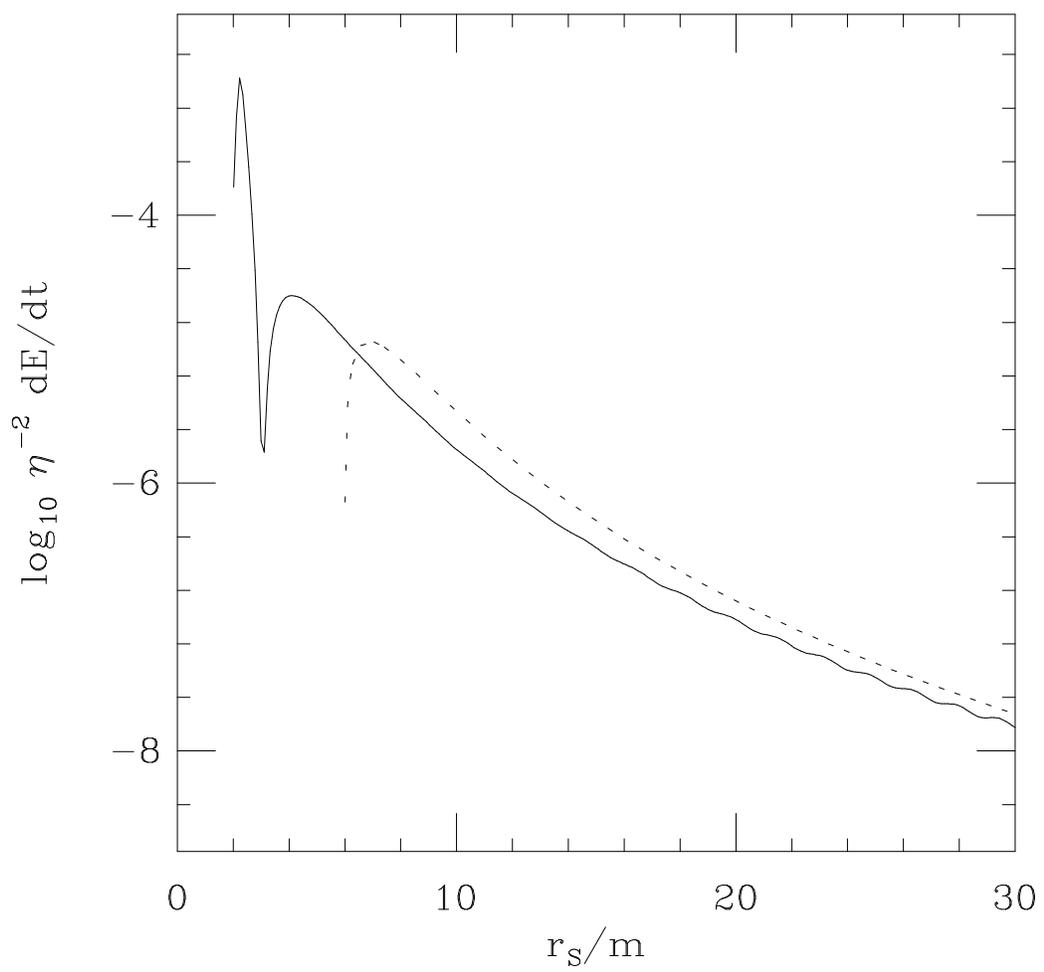
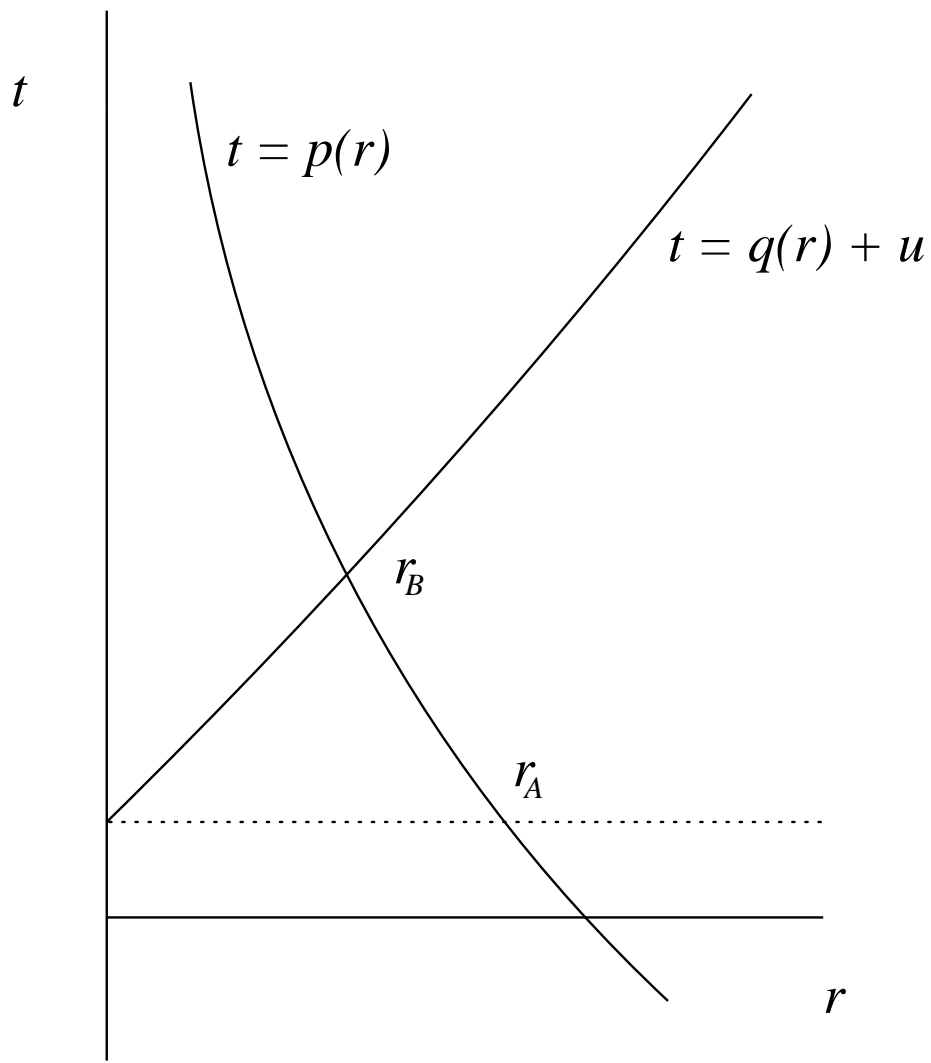
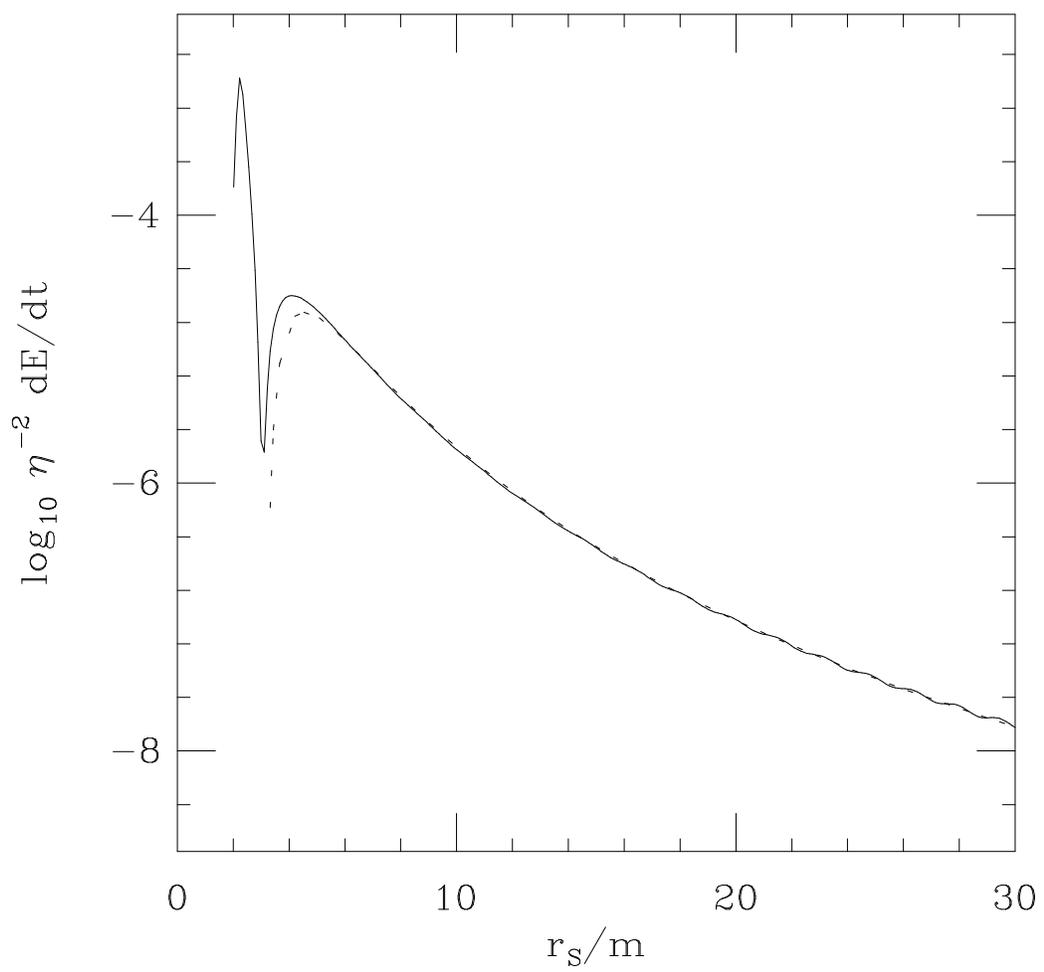
\end{document}